\documentclass[preprint,nofootinbib,aps,groupedaddress,eqsecnum]{revtex4}
\usepackage{color}
\usepackage{mathtools}
\usepackage{array}
\usepackage{booktabs} 
\usepackage{tabularx}
\usepackage{diagbox}
\usepackage{tikz}
\usetikzlibrary{decorations.pathmorphing}
\usepackage{bm}
\usepackage{amsmath,bm,amssymb,amsfonts,dcolumn,color,graphicx,latexsym,epsfig}
\usepackage{rsfso}
\usepackage[pagebackref=false, colorlinks=true]{hyperref}
\definecolor{reddish}{rgb}{0.9,0.3,0.0}  
\definecolor{blueish}{rgb}{0.1,0.1,1}
\hypersetup{
colorlinks=true,
linkcolor=reddish,
filecolor=reddish,      
urlcolor=magenta,
citecolor=blueish,
}
\usepackage{subfigure}
\usepackage{multirow}
\usepackage{natbib}
\usepackage{float}
\usepackage{booktabs}
\usepackage{pifont}
\usepackage{mathrsfs}
\usepackage{caption}
\usepackage{gensymb}
\usepackage{caption, threeparttable}
\usepackage{orcidlink}
\begin{document}

\title{Diverse regular spacetimes using a parametrised density profile}

\author{Anjan Kar\,\orcidlink{0009-0009-8623-3596}}
\email[Email address: ]{anjankar.phys@gmail.com}
\affiliation{Department of Physics, Indian Institute of Technology, Kharagpur 721 302, India}

\author{Sayan Kar\,\orcidlink{0000-0001-5854-6223}}
\email[Email address: ]{sayan@phy.iitkgp.ac.in}
\affiliation{Department of Physics, Indian Institute of Technology, Kharagpur 721 302, India}

\begin{abstract}
\noindent We explore the construction of diverse regular spacetimes (black holes and defects)
in  General Relativity (GR)
using a generic parametrised density profile (the Dekel-Zhao profile), which includes, for
specific parameter choices, various well-known examples 
usually studied in the context of dark matter halos. Our solutions, in the
Schwarzschild gauge, include new regular black holes as well as non-singular
solutions representing spacetime defects. For a sub-class of metrics, a TOV equation
approach with a chosen equation of state works. The status of the energy conditions and the issue of geodesic completeness are explored in detail. We also provide possible Lagrangian density
constructions for the matter energy-momentum tensors. Further, we study the shadow radius of the new regular black holes, and compare our findings with available observational results from the EHT collaboration.
Finally, for the defect
solution, we present a model for a stable star (a gravastar) by explicit use of
the junction conditions and obtain relevant consequences highlighting its
characteristic features. 
\end{abstract}


\maketitle

\newpage

\section{Introduction}\label{I}

\noindent It is a well-known and proven fact that General Relativity (GR) as well as other modified metric theories of gravity admit singular spacetime solutions. A singular solution may be
defined as one having (a) at least one infinite-valued curvature invariant  or (b) incomplete geodesics with or without a divergent curvature invariant. The theorems of Penrose \cite{penrose} and Hawking and Penrose \cite{hawking} which prove the existence of singularities, hold under fairly generic conditions, without
a priori assuming any theory of gravity or field equations. They
imply the inevitable presence of singularities in the sense of
geodesic incompleteness. An earlier
result due to Raychaudhuri and Komar (mentioned as the Raychaudhuri-Komar theorem in \cite{senovilla}) defined a singular spacetime using infinite curvatures and the physical property of a
divergence in the matter density/pressures. All the 
above-mentioned ways of defining a spacetime singularity are accepted by the GR community, but the
exact connections between them remains somewhat obscure since we
end up with inherently one-way statements (eg. curvature divergence implies geodesic incompleteness but not vice-versa).

\noindent A singular 
behaviour of any field (say, electromagnetic), in general, is not such a disaster
as such, but a singularity in spacetime itself (which is the gravitational field too) is indeed a major problem, since all fields live in spacetime
and may end up displaying undesirable pathologies 
while encountering the singular point or location. Thus, in a way, it
seems we will be better off if we can manage to live and work
in a nonsingular (regular) spacetime. We may be able to construct such
nonsingular spacetimes by either circumventing some of the assumptions
in the proofs of the singularity theorems \cite{senovilla} or by going over to the 
quantum regime, expecting it to `resolve' singularities in some
yet-to-be-fully-known way \cite{Horowitz, Hofmann, Casals}. We will, in this article, follow the classical route
in our quest for nonsingular (regular) spacetimes.

\noindent Currently, there exists a large variety of regular spacetime
geometries which represent regular black holes, wormholes or 
other exotic compact objects (ECO). For each of them, one can identify
a violation of one or more of the assumptions used in proving the singularity theorems in \cite{penrose, hawking}. In a recent article
\cite{borissova}, the authors have explained in detail the 
specific assumption(s) which are indeed violated for a large class of regular black holes. The question however remains whether such violations are acceptable or not. In other words, one needs to figure out if the violations lead to consequences which are dangerous and jeopardise
the very existence of these constructions. The answer to this query is
not yet completely known. We, therefore, prefer an open and liberal 
stand on this issue.

\noindent Thus, leaving aside the issue of violations of the assumptions
and their consequences, we first need to know what methods have been developed in order to construct such spacetimes. In most constructions, there is no chosen model for the matter stress energy. It has been found largely by reverse engineering, that
nonlinear electrodynamics when coupled to gravity can be a viable model
for a large variety of regular black holes \cite{Ayon1, Ayon2, Ayon3, Ayon4, Ayon5, Bronnikov1, Balart, Bronnikov3, Poshteh, Kar1, Kar2}. There are other models too, 
involving scalar fields \cite{Bronnikov2,bronnikov-rev}. But, by and large, the matter sector required for
any regular black hole to exist, is not really known or understood. 
There are alternate methods for constructing regular black holes \cite{ovalle, Bueno}, though, we are yet to have a clear algorithm with all the 
desired properties. The idea of regular primordial black holes and their relevance in cosmology has also been looked at in \cite{khlopov}.

\noindent In our work in this paper, we consider static, spherically symmetric spacetimes in the Schwarzschild gauge. Thus, we have only one metric function to worry about and we also know that the radial pressure
is always the negative of the density, i.e. $p_r=-\rho$ (assuming Einstein's equations hold, given the chosen form of the line element in the Schwarzschild gauge). We can therefore
postulate the energy density and obtain the metric function as well as
the tangential pressure $p_t$. 

\noindent How do we choose the density profile? 
For guidance, we fall back on the numerous known profiles extensively
used in studies on dark matter halos. A large number of them can be
collectively written in a single expression involving generic parameters, popularly known as the Dekel--Zhao (DZ) profile
\cite{zhao, zhao2, dekel, dz1}. 
In reality, one must note that individual profiles for dark matter distributions in galaxies 
vary widely and do not necessarily merge into a unifying profile with running parameters. Thus, the Dekel-Zhao profile merely serves as a formula for expressing some (but not all) known dark matter distributions \cite{Salucci}. Our work here, however, has little to do with dark matter and we just make use of the DZ density expression.
Special choices of the
parameters in the DZ profile lead to some of the
known regular black hole solutions in the Schwarzschild gauge.
Newer solutions in the Schwarzschild gauge have been found in \cite{dz2, Sekhmani2}.
A recent example of a singular black hole solution with $-g_{00} \neq (g_{11})^{-1}$ (i.e. not in the Schwarzschild gauge) has also been 
reported in \cite{ovgun}.

\noindent However, there still exist possibilities of constructing
newer spacetimes which have not been obtained or looked at before. This includes new regular black holes and another class of regular spacetimes where a solid angle deficit arises as a key feature. 
This latter class
(i.e. those with a solid angle deficit) belong to the
wider category commonly known as `spacetimes with defects'.
Prominent examples include the global monopole and the cosmic string (angle deficit)  \cite{Barriola, Nucamendi, Marunovic, Carames, Vilenkin}. Thus, our aim here is to focus on constructing novel examples of these
two types of spacetimes using the chosen parametrised density profile.   
We also check their geometric features and nonsingular character (curvature invariants as well as geodesic completeness).
Further, we note that instead of postulating the energy density, one may propose an equation of state and use the Tolman-Oppenheimer-Volkoff (TOV) equation to
find the metric function. Though this approach has been analysed
recently \cite{Luongo}, we revisit some features around this method, briefly.
In order to connect our work with some observations, we find the shadow cast by the new regular black hole and constrain the metric parameters through available EHT observations.
Finally, we use the regular spacetime representing a defect, to model the interior of a star
and apply the junction conditions at the boundary, using a Schwarzschild exterior. A thin shell construction is required to build a satisfactory model. The stability of this model star is thereafter investigated in some detail.

\noindent Thus, through our work here, we intend to illustrate 
and analyse the diversity in the spectrum of possibilities 
(representing regular spacetimes) which
emerge from the choice of a parametrised density profile. 

\noindent Our article is organised as follows. In the next section, we
discuss the parametrised Dekel-Zhao density profile and give an outline of our
program of constructing solutions. In Section \ref{III}, we analyse the
new regular black holes and study their shadow radius.
Section \ref{IV} presents a new regular solution which has a solid angle deficit
and represents a defect. Our model for a stable star using the defect
solution is spelt out in \ref{VI}, where we also elaborate on the
junction conditions and the need of a thin shell at the boundary.  In Section \ref{V}, we briefly present the
TOV equation approach.
Section \ref{VII} contains our conclusions with some pointers towards future work. 

\noindent It is important to mention that we work in geometrical units, i.e. $G=c=1$. However, the SI unit is followed when referring to some observations.


\section{The parametrised Dekel-Zhao density profile and our approach}\label{II}
\noindent 
As mentioned in the Introduction, we will work in the framework of
GR assuming a static, spherically symmetric Schwarzschild gauge metric ansatz. Our line element is assumed to be:
\begin{equation}\label{2.3}
    ds^{2}=-f(r)dt^{2}+\frac{dr^{2}}{f(r)}+r^{2}\left(d\theta^{2}+\sin^{2}{\theta} d\phi^{2}\right)
\end{equation}
where $f(r)=1-\frac{2m(r)}{r}$, and $m(r)$ is the mass function.
The Einstein field equations are given, in general, as:
\begin{equation}\label{2.1}
  G_{\mu\nu}= R_{\mu\nu}-\frac{1}{2}g_{\mu\nu}\texttt{R}=8\pi T_{\mu\nu}
\end{equation}
where $G_{\mu\nu}$ is the Einstein tensor, $R_{\mu\nu}$ the Ricci tensor, $g_{\mu\nu}$ 
the spacetime metric and $\texttt{R}$ the Ricci scalar.
We further assume the energy-momentum tensor $T_{\mu\nu}$ in the following form,
\begin{gather}\label{2.2}
 T^{\mu}_{\nu}
 =
  \begin{pmatrix}
   -\rho & 0 & 0&0 \\
   0 & p_r & 0 &0\\
   0 & 0 & p_t&0\\
   0&0&0&p_t\\
   \end{pmatrix}
\end{gather}
where $\rho$ is the energy density, $p_r$ the radial pressure and $p_t$ the tangential pressure--all defined in the frame basis. We now obtain the Einstein tensor components in the frame basis, for the given metric
and equate it to the energy-momentum tensor components. 
The choice of the line element in Eq.(\ref{2.3}), in the Schwarzschild gauge, automatically leads to the equation $G_{rr} = -G_{tt}$, where $G_{ij}$ is the 
Einstein tensor in the frame basis. This results in
the following simple equations arising from the Einstein equations $G_{ij} = 8\pi T_{ij}$:
\begin{equation}\label{2.4}
    \rho=-p_r=\frac{2m^{\prime}}{8\pi r^2}, \hspace{1cm} p_t=-\frac{m^{\prime\prime}}{8\pi r}
\end{equation}
where a prime denotes the radial derivative. Note that 
for the chosen spacetime, a supporting matter energy-momentum tensor
must have the following properties: (a) it has non-zero radial and tangential pressures as well as a non-zero energy density and (b) the radial pressure is negative and equal to the energy density.

\noindent Thus, for a chosen $\rho$, one can easily evaluate the mass function $m(r)$ and, consequently, the radial and tangential pressures from the above equations (\ref{2.4}). 

\noindent We choose the parametrised Dekel-Zhao density profile \cite{zhao}
given as,
\begin{equation}\label{2.5}
    \rho=\frac{\rho_{0}\left(\frac{r}{R}\right)^{\mu-3}}{\left(1+\left(\frac{r}{R}\right)^{\nu}\right)^{\frac{\mu+\alpha}{\nu}}}
\end{equation}
where the parameters $\mu$, $\nu$, $\alpha$ are dimensionless, $R$ 
has length dimensions and may be called the length scale parameter. When $\mu=3$, $\rho_{0}$ represents the central density.
One can easily identify the above generic density distribution with known dark matter halo density profiles for specific parameter values. A partial list is shown in Table \ref{table1}. 
\begin{table}[h]
\centering
\begin{tabular}{ |c|c|c| } 
\hline
{\bf Parameter values} & {\bf Density distribution} & {\bf Known dark matter profile} \\
\hline
$\mu=2, \nu=1, \alpha=0$ &  $\frac{\rho_{0}R}{r}\left(1+\frac{r}{R}\right)^{-2}$ & \text{NFW profile}\cite{NFW}\\
\hline
$\mu=3, \nu=2, \alpha=0$ & $\rho_{0}\left(1+\frac{r^2}{R^2}\right)^{-3/2} $ & \text{King profile}\cite{King}\\ 
\hline
$\mu=3, \nu=2, \alpha=-1$ & $\rho_{0}\left(1+\frac{r^2}{R^2}\right)^{-1}$ &  \text{Pseudo Isothermal profile}\cite{Isothermal} \\
\hline
$\mu=3, \nu=2, \alpha=2$ & $\rho_{0}\left(1+\frac{r^2}{R^2}\right)^{-5/2}$ & \text{Plummer profile}\cite{Plummer}\\
\hline
$ \mu=2, \nu=1, \alpha=1$ &  $\frac{\rho_{0}R}{r}\left(1+\frac{r}{R}\right)^{-3}$ & \text{Hernquist profile}\cite{Hernquist}\\
\hline
\end{tabular}
\caption{Partial list of some well-known dark matter density profiles 
included in the Dekel-Zhao profile}
\label{table1}
\end{table}
\noindent Our aim here is to use the DZ density profile with chosen parameters and first find $m(r)$ and subsequently,
$p_r(r)$ and $p_t(r)$. We do not claim any direct relation of our 
work with cold dark matter which, as is well-known, has negligible pressures.

\noindent It is not a necessity that the GR spacetimes sourced by the above density profiles will always lead to a regular geometry for any
choice of parameters. Our purpose here is to
see if it does, for some chosen parameter values. To ensure that there is no singularity at the origin, we have to check the behaviour of all independent curvature scalars, namely the  Ricci scalar ($g_{\mu\nu}R^{\mu\nu}$), Ricci contraction ($R_{\mu\nu}R^{\mu\nu}$) and Kretschmann scalar ($R_{\mu\nu\lambda\delta}R^{\mu\nu\lambda\delta}$) \cite{Ayon1, Zakhary, Hu}. The Ricci scalar and the Ricci contraction for the spacetime in Eq.(\ref{2.3}) can be written in terms of the energy density and its radial derivative as follows,
\begin{equation}\label{2.6}
    g_{\mu\nu}R^{\mu\nu}=8\pi (4\rho +r \rho^{\prime}),\hspace{1cm} R_{\mu\nu}R^{\mu\nu}=32 \pi^2\left(8\rho^2+4r\rho\rho^{\prime}+r^2\rho^{\prime 2}\right)
\end{equation}
Regularity of the above scalars demand $\rho$ and $\rho^{\prime}$ to have a finite value as $r\to 0$. As a result, we have the following restrictions on the density parameters: $\mu\geq 3$, $\nu>0$ and $\alpha>-3$. Now, the third independent scalar, the Kretschmann scalar, cannot be fully described in terms of $\rho$. It is found that the Kretschmann scalar depends on the metric function ($m(r)$) along with $\rho$ and $\rho^{\prime}$.
\begin{equation}
  R_{\mu\nu\lambda\delta}R^{\mu\nu\lambda\delta}=\frac{48 m^2}{r^6}+\frac{64\pi m}{r^3}(-2\rho+r\rho^{\prime})+64\pi^2(4\rho^2+r^2\rho^{\prime 2})  
\end{equation}
Therefore, to comment on the regularity of the Kretschmann scalar, it is essential to look at each individual case, as defined
for chosen parameter values. We will now work out two specific examples
which are, as yet, not studied in the literature.


\section{Regular black holes}\label{III}
\subsection{A new regular black hole and its properties}
\noindent We first consider the King dark matter density profile \cite{King}, which obeys the conditions for a regular Ricci scalar and Ricci contraction (i.e. $\mu\geq 3$, $\nu>0$, $\alpha>-3$). For $\mu=3, \nu=2,$ and $ \alpha=0$ the King density profile is written as,
\begin{equation}\label{3.1}
    \rho(r)=\frac{\rho_{0}}{\left(1+\frac{r^2}{R^2}\right)^{3/2}}
\end{equation}
Solving the Eq.(\ref{2.4}) with this choice for $\rho$, one 
obtains the following metric function
\begin{equation}\label{3.2}
    f(r)=1-\frac{2M}{r}+\frac{8\pi\rho_{0}R^{3}}{\sqrt{r^2+R^2}}+\frac{8\pi\rho_{0}R^3}{r}\ln{\left(\frac{\sqrt{r^2+R^2}-r}{R}\right)}
\end{equation}
Note that the above solution reduces to the Schwarzschild solution when $\rho_{0}R^2=0$. We claim that it represents a family of regular black holes for $M=0$. For nonzero $M$, we get a singular solution. This is verified by deriving and analysing the curvature
invariants (see the next subsection).  

\noindent It is interesting to note that the above metric function can be expanded asymptotically as a combination of powers of $r^{-1}$ and positive powers of $\ln{r}$, i.e. as $r\to\infty$
\begin{equation}
    f(r)=1+8\pi\rho_{0}R^2\left[\frac{R}{r}-\frac{\ln{(2r/R)}}{r/R}\right]+O\left(\frac{1}{r^3}\right)
\end{equation}
In the literature, spacetimes with such asymptotic behaviour are known as polyhomogeneous spacetimes \cite{Chrusciel}. This class of spacetimes is claimed to be more realistic than the class of spacetimes which admit a smooth, flat expansion \cite{Godazgar, He}.  

\noindent It is also evident from the above expansion that the metric in Eq.(\ref{3.2}) is asymptotically flat.
Moreover, at small values of $r$, the metric function 
reduces to that for de-Sitter space, i.e.
\begin{equation*}
    f(r)\to 1-c_1^2r^2, \hspace{1cm} \text{as}\hspace{1cm} r\to 0
\end{equation*}
To understand the spacetime structure, we examine the roots of the equation $g_{tt}=0$, which represent the horizons in the above geometry 
\cite{vishu, mt}. 
\begin{figure}[h]
\centering
\includegraphics[width=0.6\textwidth]{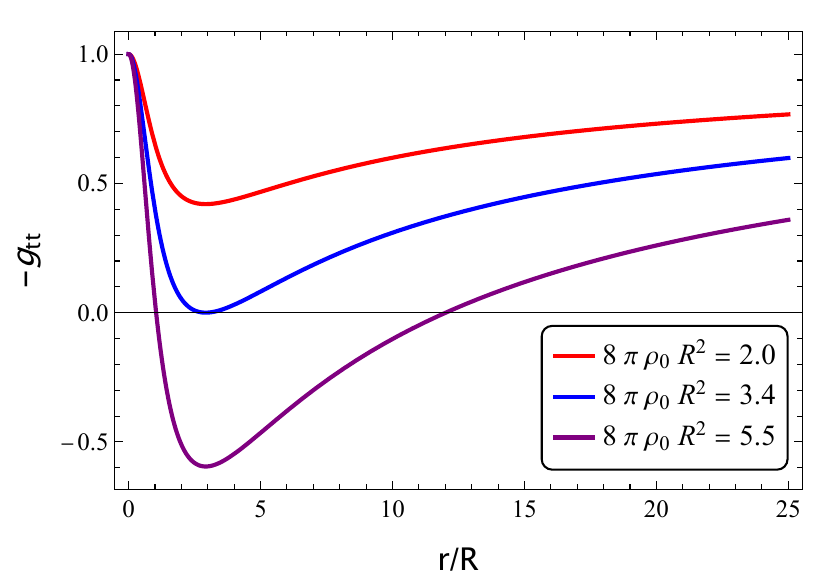}
\caption{Graph of the redshift function with $r/R$ for various parameter values.}
\label{fig:Horizon_RBH}
\end{figure}
Based on the number of real, positive roots of the horizon equation and Figure \ref{fig:Horizon_RBH}, one can infer that the above geometry represents a family of black holes with double horizon (inner horizon and outer horizon) when $8\pi\rho_{0}R^2>3.448$. A single horizon black hole exists for $8\pi\rho_{0}R^2=3.448$. Horizon-less compact objects are obtained when $8\pi\rho_{0}R^2<3.448$ (no real positive root of the horizon equation $g_{tt}=0$).

\subsubsection{Regularity of curvature scalars and geodesic completeness}
\noindent To verify that the metric in Eq.(\ref{3.2}) represents a regular spacetime, we examine the three independent curvature scalars explicitly. A smooth and continuous behaviour of the scalars over the entire domain of the radial coordinate is a necessary condition to prove that the metric is genuinely regular. The regularity of the Ricci scalar and Ricci contraction are expected trivially from our parameter choices.

\noindent We have, for the Ricci scalar,
\begin{equation}
    g_{\mu\nu}R^{\mu\nu}=\frac{8\pi\rho_{0}R^3(r^2+4R^2)}{(r^2+R^2)^{5/2}}, \hspace{1cm}
\end{equation}
and the Ricci contraction is given by,
\begin{equation}
    R_{\mu\nu}R^{\mu\nu}=\frac{32\pi^2\rho_{0}^{2}R^6(5r^4+4r^2R^2+8R^4)}{(r^2+R^2)^5}
\end{equation}
Obviously, they are finite everywhere.
We do not write the full expression for the Kretschmann scalar but
quote its $r\rightarrow 0$ value. This is given as,
\begin{equation}
    \lim_{r\to 0}R_{\mu\nu\lambda\delta}R^{\mu\nu\lambda\delta}=\frac{512\pi^2\rho_{0}^2}{3}
\end{equation}
Thus, the finiteness of the three independent curvature scalars as $r\rightarrow 0$ indicates that there is no curvature singularity. However, the regularity of curvature invariants is not sufficient to conclude that the spacetime is non-singular in the extended domain of coordinates \cite{Hawking, Wald}. According to \cite{Modesto, Carballo1, Carballo2}, completeness of all causal geodesics is a necessary requirement for the regularity of spacetime. This is because 
a singularity may be defined only in the sense of geodesic incompleteness.
Let us see how one can check geodesic completeness for our spacetime.

\noindent If the affine parameter of a causal geodesic to reach $r=0$ is finite, in a given geometry, one can `mathematically' extend the geodesic further to negative values of $r$. As coordinates
and coordinate systems by themselves are not physical quantities, such an extension is mathematically possible. 
A spacetime is geodesically complete when the causal geodesics are
extendible and well defined in the negative values of $r$ (just as they are for positive $r$), and the extension is
valid right up to $r\to -\infty$.
It can be therefore be stated that, in a geodesically complete spacetime, the affine parameter varies from $-\infty$ to $+\infty$. One may consult \cite{Modesto} for further details on this approach. 
 
\begin{figure}[h]
\centering
\includegraphics[width=0.6\textwidth]{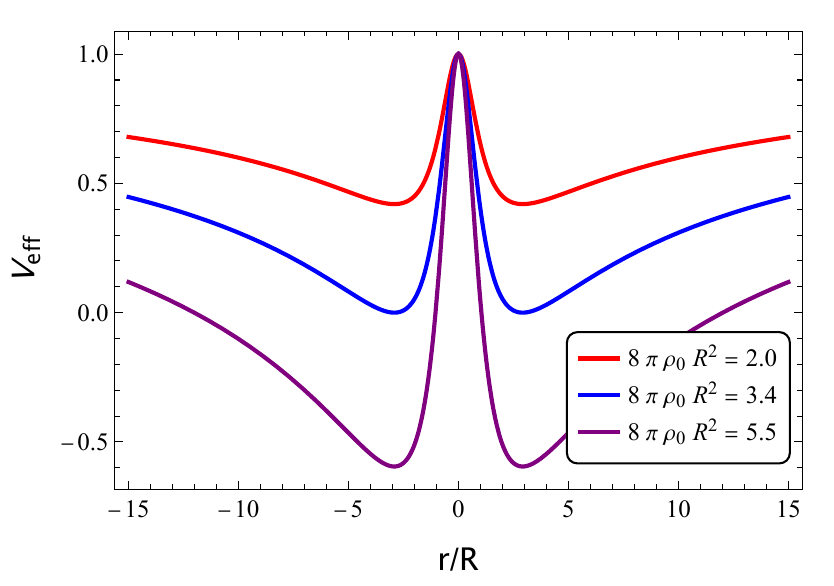}
\caption{Plot of the effective potential for radial timelike geodesic in the extended domain of $r$, for different parameter values.}
\label{fig:ep_RBH}
\end{figure}

\noindent For radial timelike geodesics, geodesic completeness can be shown by examining the smooth and continuous behaviour of the effective potential ($V_{eff}=E^2-\dot{r}^2$) in the extended domain of $r$ \cite{Modesto}. Here, overdot represents the derivative with respect to the affine parameter and the conserved quantity $E=-g_{tt}\dot{t}$. For our geometry in Eq.(\ref{3.2}), the effective potential for radial timelike geodesics is expressed as
\begin{equation}\label{3.8}
    V_{eff}=-g_{tt}=1+\frac{8\pi\rho_{0}R^{3}}{\sqrt{r^2+R^2}}+\frac{8\pi\rho_{0}R^3}{r}\ln{\left(\frac{\sqrt{r^2+R^2}-r}{R}\right)}
\end{equation}
In Figure \ref{fig:ep_RBH}, we demonstrate the effective potential in the extended domain of the radial coordinate for different parameter values. Its smooth and continuous behaviour confirms the completeness of timelike geodesics.
Similar to timelike geodesics, geodesic completeness can be shown for null geodesics as well, for this spacetime. Thus, the regular nature of the geometry in Eq.(\ref{3.2}) is confirmed through scalar curvatures as well as from the behaviour of causal geodesics.

\subsubsection{Energy conditions}
\begin{figure}[h]
\centering
\includegraphics[width=0.6\textwidth]{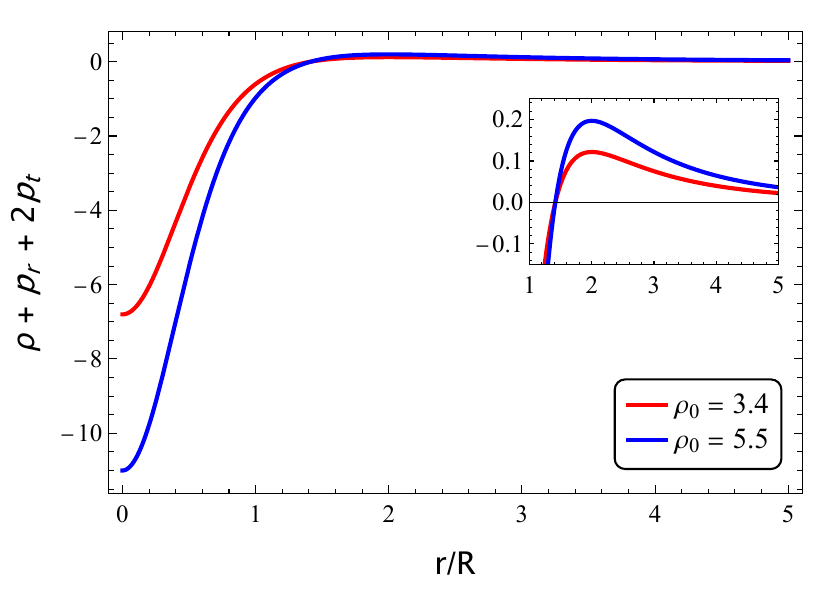}
\caption{Plot of the SEC with $r/R$ for different values of $\rho_{0}$.}
\label{fig:SECking}
\end{figure}

\noindent Let us now examine the different energy conditions for the 
matter required to support such a regular geometry. The energy density ($\rho$) is already assumed in Eq.(\ref{3.1}). The other components of the energy-momentum tensor, as obtained assuming Einstein equations
of GR hold, are the following,
\begin{equation}\label{3.9}
    p_r=-\rho, \hspace{1cm}p_t=\frac{\rho_{0}R^3(r^2-2R^2)}{2(r^2+R^2)^{5/2}}
\end{equation}
It is evident from the above expressions (and the chosen $\rho$) that
\begin{align}
    \rho>0, \hspace{1cm} \rho+p_r=0, \hspace{1cm} \rho+p_t=\frac{3\rho_{0}R^{3}r^2}{2(r^2+R^2)^{5/2}}>0
\end{align}
\begin{figure}[h]
\centering
\includegraphics[width=0.6\textwidth]{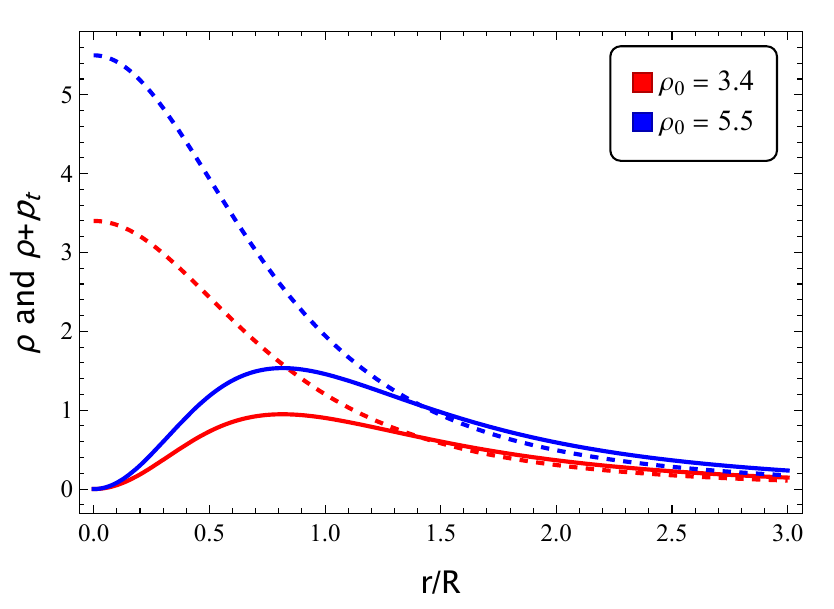}
\caption{Plot of $\rho$ and $\rho + p_t$ with $r/R$ for different values of $\rho_{0}$. The dashed lines and solid lines represent $\rho$ and $\rho + p_t$, respectively. The same coloured lines have equal $\rho_{0}$.}
\label{fig:NECking}
\end{figure}
\noindent Therefore, the required matter obeys the Null Energy Condition (NEC) and Weak Energy Condition (WEC)  over the entire domain of the radial coordinate. This is also confirmed by Figure \ref{fig:NECking}. 
For the Strong Energy Condition (SEC) to hold, $\rho+p_r+2p_t$ must be non-negative. We have 
\begin{equation}
    \rho+p_r+2p_t=\frac{\rho_{0}R^3(r^2-2R^2)}{(r^2+R^2)^{5/2}}
\end{equation}
Thus, SEC is violated for $r<\sqrt{2}R$, which is also evident from Figure \ref{fig:SECking}. When $r$ is smaller than $\sqrt{2}R$, gravitational attraction
is absent and we have, instead, a repulsion, which is one of the necessary conditions to form a regular centre \cite{Ansoldi, Zaslavskii}.

\subsubsection{Matter source for the geometry}
\noindent We mentioned in the Introduction that nonlinear electrodynamics minimally coupled to gravity can source several regular black holes \cite{Ayon1, Ayon2, Ayon3, Ayon4, Ayon5, Bronnikov1, Balart}. Similarly, we may model the required matter for the above-discussed regular black hole in terms of a magnetic monopole governed by a specific nonlinear electrodynamics.
We have found that the following nonlinear electrodynamics Lagrangian density minimally coupled to GR, can be used to model the required matter energy-momentum tensor for our regular black hole. We have the following expression for $\mathcal{L}(F)$,
\begin{equation}
    \mathcal{L}(F)=\frac{\delta (2F)^{3/4}}{(1+\gamma\sqrt{2F})^{3/2}}
\end{equation}
where $F=\frac{1}{4}F_{\mu\nu}F^{\mu\nu}$, $\delta=-\frac{\rho_{0}R^3}{q_{m}^{3/2}}$ and $\gamma=\frac{R^2}{q_{m}}$. The only nonzero component of the field strength tensor ($F_{\mu\nu}$) is $F_{\theta\phi}$, making the regular black hole a purely magnetic solution. The magnetic source is identified with a magnetic monopole having radial magnetic field $B_{r}=\frac{q_{m}}{r^2}$ and $q_{m}$ may be understood as total magnetic charge. Note that the details of the derivation of the above Lagrangian density are not shown here. One can go through \cite{Kar1, Kar2} 
to learn more about the derivation. It is useful to note that even though we have used the King density profile (commonly used in studies on dark matter) we are simultaneously able to construct a nonlinear electrodynamics model for the source. This, as we shall see later, is similar to the Bardeen case which has a Plummer density profile but
a source (with pressures) which relates to a nonlinear electrodynamics model.

\subsubsection{Shadow radius and EHT observation}
\noindent Let us now study null geodesics in the new regular black hole geometry and compute the shadow radius in the equatorial plane. The formation of black hole shadow can be understood as the result of interaction between the strong gravitational field caused by the black hole and the surrounding light rays. Photons interact with the gravitational field of a black hole, and are either trapped or they escape from it in accordance with their impact parameter $(L/E)$, where $L$ and $E$ are defined later below. As a consequence, a distant observer perceives a dark region in the sky, which is known as the black hole shadow. To find the boundary of the dark region in the observers sky, we have to evaluate the critical photon orbit, which differentiates between the scattered and trapped photons. This is done by analysing the null geodesics, which, in the equatorial plane, are obtained from
\begin{equation}
    \dot{r}^2+V(r)=0
\end{equation}
where $V(r)=-E^2+\frac{L^2}{r^2}f(r)$. The conserved quantities are $E=-g_{tt}\dot{t}$ and $L=r^2\dot{\phi}$. The critical photon orbit radius ($r_c$) known as the radius of the `photon sphere' can be evaluated from the potential $V(r)$. The critical orbit satisfies the following conditions:
\begin{equation}
    V(r_{c})=0, \hspace{1cm} \frac{\partial V}{\partial r}\bigg|_{r=r_{c}}=0, \hspace{1cm} \frac{\partial^2 V}{\partial r^2}\bigg|_{r=r_{c}}<0
\end{equation}
For $8\pi\rho_{0}R^2\neq 0$, the equation with a first derivative of $V$ has two real positive roots. By performing the stability analysis of the potential $V(r)$, one may determine the critical root which represents the radius of the unstable circular orbit or the photon sphere radius \cite{Virbhadra}.
\begin{figure}[h]
\centering
\includegraphics[width=0.6\textwidth]{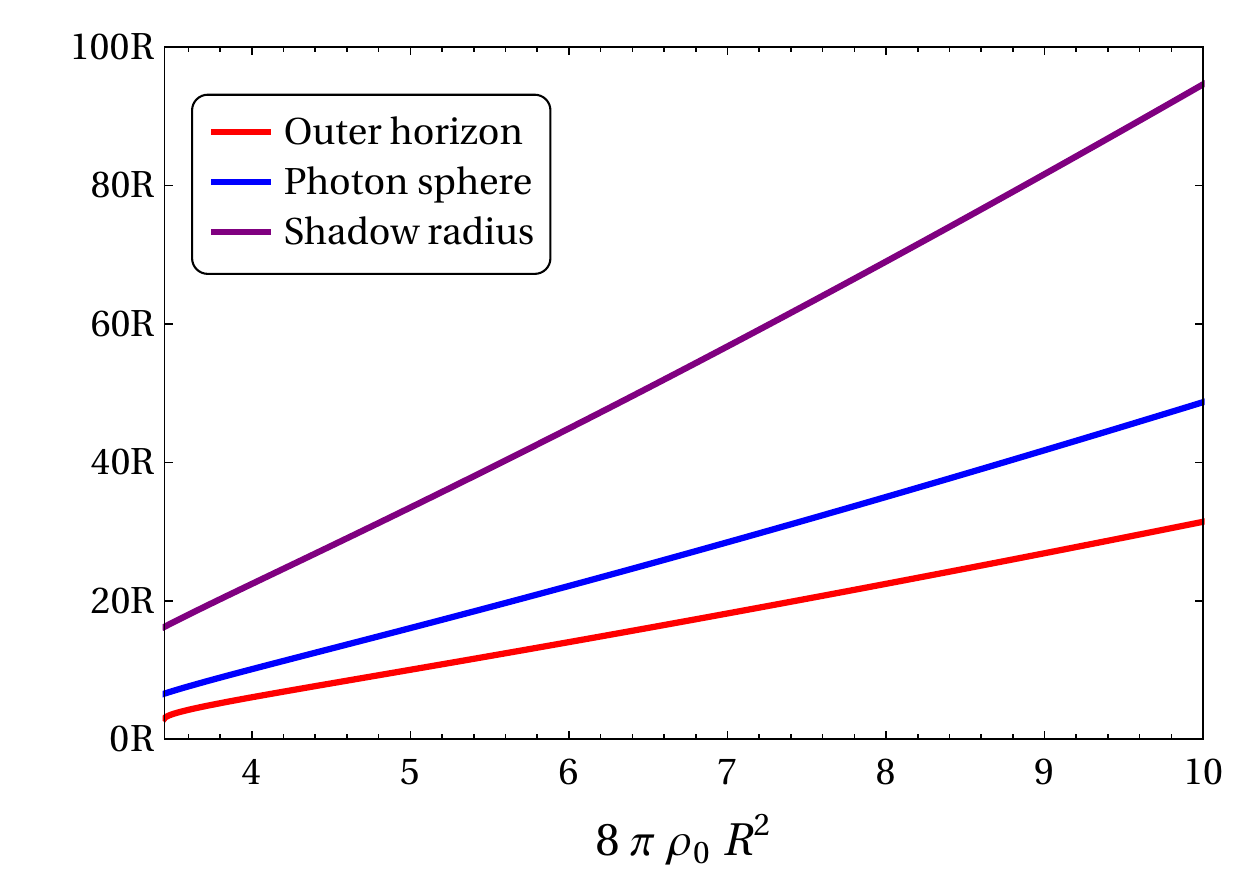}
\caption{Plot depicting the shadow radius (purple line), photon sphere radius (blue line) and outer horizon radius (red line) as a function of $8\pi\rho_{0}R^2$. The parameter $8\pi\rho_{0}R^2$ on the $x$-axis begins at $3.45$ where the outer and inner horizons coincide. When $8\pi\rho_{0}R^2\lessapprox3.45$, there is no horizon. However, the photon sphere radius extends to $8\pi\rho_{0}R^2\approx3.25$ (not included in the plot).}
\label{fig:Shadow}
\end{figure}
In Figure \ref{fig:Shadow}, we demonstrate the photon sphere radius (blue line) as a function of $8\pi\rho_{0}R^2$, which is larger than the outer horizon radius (red line). The shadow profile of a static, spherically symmetric geometry as seen in the distant observers sky is circular. Its radius can be calculated in terms of celestial coordinates \cite{bardeenshadow}. The shadow radius is directly related to the photon sphere radius and is expressed as \cite{Volker, Chen},
\begin{equation}
    r_{sh}=\frac{r_c}{\sqrt{f(r_c)}}
\end{equation}
In Figure \ref{fig:Shadow}, we show the variation of the shadow radius (purple line) with the metric parameter $8\pi\rho_0 R^2$. 
We note that $r_{sh}$ is larger than the photon sphere radius. Although the photon sphere radius represents the boundary between the infalling photons and the scattered ones, the distant observer sees its projection in his/her sky. Thus, $r_{sh}>r_c$. 

\begin{figure}[h]
\includegraphics[width=0.45\textwidth]{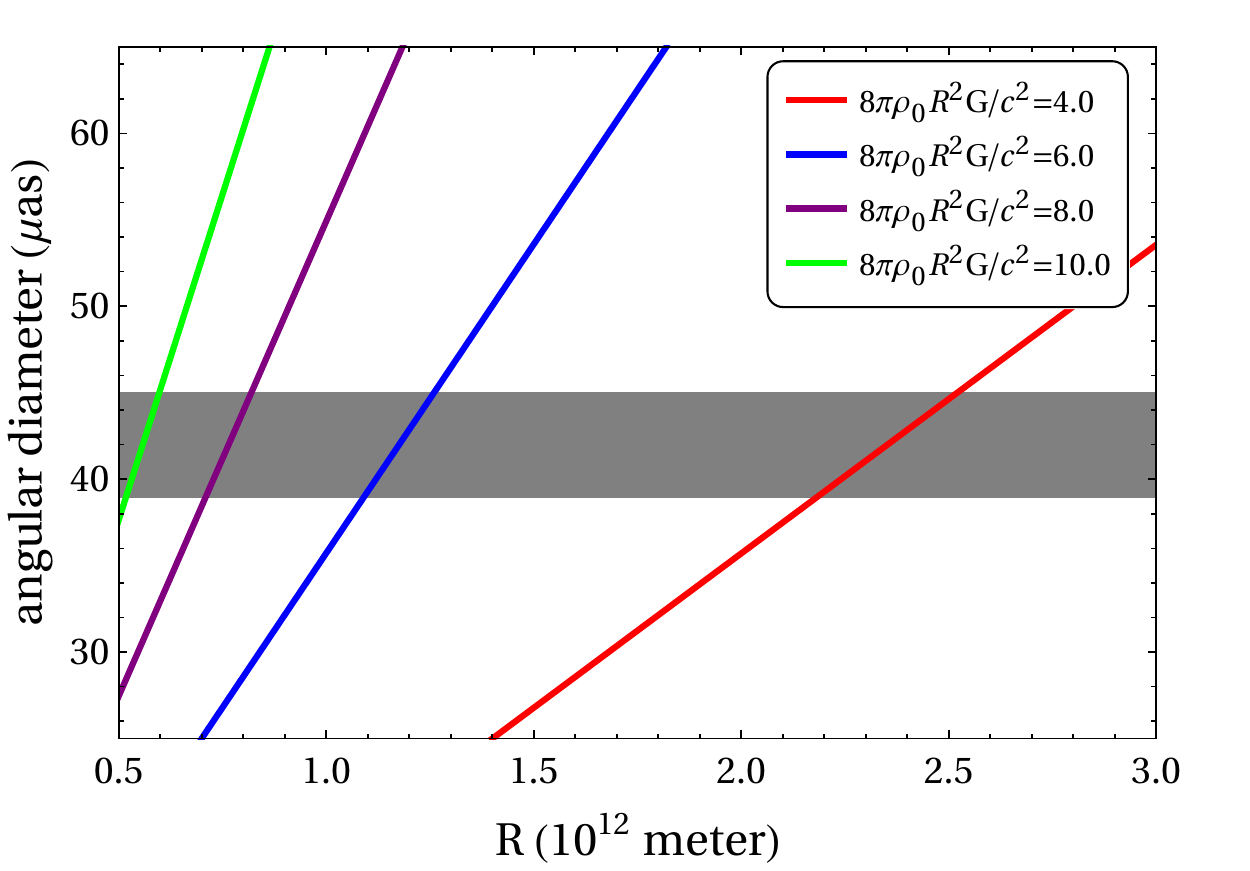}
\includegraphics[width=0.45\textwidth]{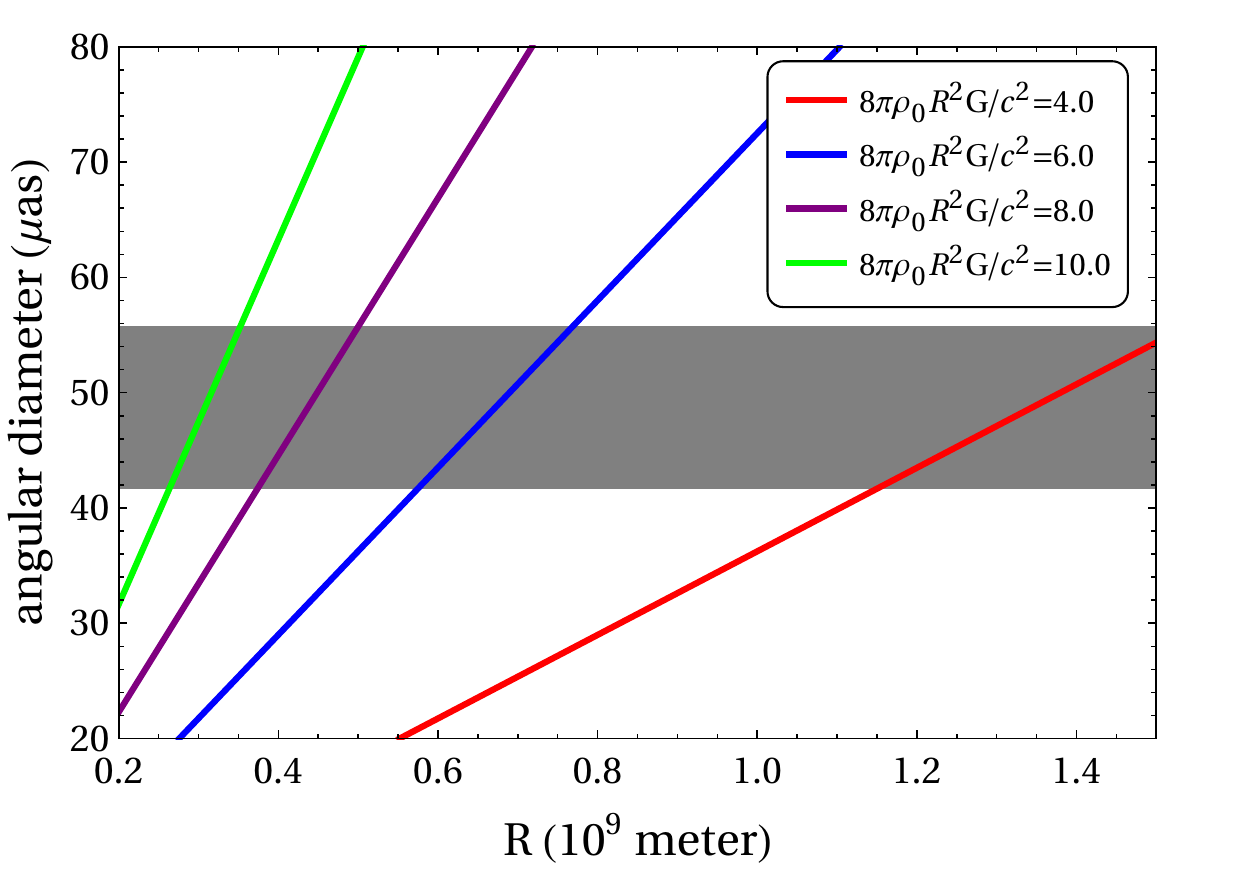}
\caption{The different coloured lines represent the angular diameter of the theoretical shadow as a function of $R$ for different $8\pi\rho_{0}R^2G/c^2$. The dark grey regions denote the observed angular diameter of the respective compact objects, considering the error bar. The range of $R$ which lies in the dark region, for a certain $\rho_{0}$, matches with observations. (Left plot) Parameters are constrained using the shadow of M87$^*$ where $\Phi=(42\pm 3)\mu as$ is taken into account. (Right plot) Parameters are constrained utilising the shadow of SgrA$^*$ where $\Phi=(48.7\pm 7)\mu as$ is considered.}
\label{fig:shadowmatch}
\end{figure}

\noindent The circular shadow profile may be used to estimate the metric parameters by comparing it with EHT observational results. The EHT observations suggest that the shadow profiles of the compact objects are not completely circular. However, a preliminary and qualitative estimation of metric parameters is possible due to the feature that the deviations from circularity of the observed shadow profiles are indeed small. 
We choose some specific values of the metric parameter $8\pi\rho_{0}R^2$ and calculate the corresponding theoretical angular diameter of the shadow in terms of $R$ (using SI units).
By comparing it with EHT data, the possible orders of $R$ and $\rho_0$ are obtained. According to EHT observations, the angular diameter of black hole M87$^{*}$ is $\Phi=(42\pm 3)\mu as$ \cite{Akiyama1, Akiyama5, Akiyama6}. A later analysis of the same data has reported the angular diameter to be $\Phi=41.5\pm 0.6\mu as$ \cite{Medeiros}. It is located at a distance of $(16.8\pm 0.8) Mpc$ \cite{Blakeslee,Bird, Cantiello}. In Figure \ref{fig:shadowmatch} (left), we demonstrate the range of $R$ with some specific values of $8\pi\rho_{0}R^2$, which tallies with the angular diameter of the shadow of M87$^*$ (we consider $\Phi=(42\pm 3)\mu as$).
It is found that for the chosen $8\pi\rho_{0}R^2$, the length scale parameter ($R$) and the central density ($\rho_{0}$) are the order of $10^{12}$ meter and $10^2$ kg/m$^3$ respectively.
Similarly, the angular diameter of the shadow for SgrA$^*$ is reported as $(48.7\pm 7) \mu as$ \cite{Akiyama12, Akiyama17}. There are several distance measurements of SgrA$^*$ \cite{Do, Gravity1, Gravity2}. For our study, we consider the distance measurement of the {\em Gravity collaboration}, which is $(8277\pm 9 \pm33)pc$ by accounting for the optical aberrations \cite{Gravity1, Gravity2}.
Figure \ref{fig:shadowmatch} (right) shows the estimated values of the length scale parameter $R$ and $\rho_0$ of the order of $10^{9}$ meter and $10^8$ kg/m$^3$ respectively for the chosen $8\pi\rho_0R^2$. Thus, we may summarise that the above-discussed regular black hole is a viable solution as far as the observed shadow of compact objects is concerned.
The values of $\rho_0$ and $R$ mentioned above, however, have no real link with cold dark matter in galaxies. 

\noindent Apart from  the above example of a regular black hole  
constructed using the King profile, one may choose other values of the
parameters in the density profile to construct newer regular black holes. 
Following the analysis stated above one may explore their properties in a
similar way.

\subsection{The known regular black hole solutions}
\noindent In this subsection, we identify some of the known regular black hole solutions in the literature which may be obtained by choosing 
the values of the parameters in the Dekel-Zhao density profile. It is easy to check the following facts.

\noindent (i) When $\mu=3$, $\nu=2$ and $\alpha=1$, the density distribution becomes
\begin{equation}
    \rho(r)=\frac{\rho_{0}}{\left(1+\frac{r^2}{R^2}\right)^2}
\end{equation}
Solving Eq.(\ref{2.4}), the corresponding metric function is
\begin{equation}
    f(r)=1-8\pi R^2\rho_{0}\frac{\tan^{-1}(r/R)}{2r/R}+\frac{8\pi\rho_{0}R^{4}}{2(r^2+R^2)} 
\end{equation}
The above metric function may be recognised as the Dymnikova regular metric \cite{Dymnikova}.

\noindent (ii) For $\mu=\nu=3$, $\alpha=1$ we have the following density profile,
\begin{equation}
    \rho=\frac{\rho_0}{\left(1+\left(\frac{r}{R}\right)^3\right)^{\frac{4}{3}}}
\end{equation}
The corresponding metric function can be expressed as,
\begin{equation}
    f(r)=1-\frac{8\pi\rho_{0}R^3}{r}\left\{1-\left(1+\left(\frac{r}{R}\right)^{3}\right)^{-\frac{1}{3}}\right\}
\end{equation}
This spacetime is identified with the Bronnikov regular metric \cite{Bronnikov1}.

\noindent(iii) For, $\nu=\alpha$ and $\mu\geq 3$, the density distribution becomes,
\begin{equation}
    \rho=\frac{\rho_{0}\left(\frac{r}{R}\right)^{\mu-3}}{\left(1+\left(\frac{r}{R}\right)^{\nu}\right)^{\frac{\mu+\nu}{\nu}}}
\end{equation}
This leads to the known generalised regular black hole solutions mentioned in the literature multiple times \cite{Fan, Neves},
\begin{equation}
    f(r)=1-\frac{8\pi\rho_{0}R^2(r/R)^{\mu-1}}{\mu(1+(r/R)^{\nu})^{\frac{\mu}{\nu}}}
\end{equation}
Among further special cases are the popular regular black hole solutions.
We have: (a) $\mu=3$, $\alpha=\nu=2$ as the Bardeen solution \cite{Bardeen}, (b) $\mu=\alpha=\nu=3$ is the Hayward solution \cite{Hayward}. It is interesting to note, that the density profile corresponding to the Bardeen solution is identified with the Plummer dark matter density distribution \cite{Plummer}.


\section{Regular spacetimes with a defect}\label{IV}
\noindent This section focuses on a specific subclass of the parametrised Dekel-Zhao density profile, which can source regular spacetime solutions with a defect, in particular solutions with a solid angle deficit (see discussion below for explicit definitions). In the literature, there are several solutions
(singular as well as regular) which exhibit defects, including the well-known global monopole spacetime \cite{Barriola, Nucamendi, Marunovic, Carames} and cosmic strings \cite{Vilenkin}. Aligning with the general theme of this article, i.e. the search for regular solutions, we will focus
on spacetimes without singularities.
In our case, the specific defect subclass can be obtained when
\begin{equation}\label{4.1}
    \mu+\alpha=\nu, \hspace{1cm} \alpha=-1
\end{equation}
The parametrised density profile for the above conditions reduces to,
\begin{equation}\label{4.2}
    \rho=\frac{\rho_{0}\left(\frac{r}{R}\right)^{\nu-2}}{1+\left(\frac{r}{R}\right)^{\nu}}
\end{equation}
By solving the Einstein equations (\ref{2.4}), the metric function ($f(r)$) corresponding to the above density profile 
is found to be,
\begin{equation}\label{4.4}
    f(r)=1-8\pi\rho_{0}R^2+8\pi\rho_{0}R^2\hspace{0.1cm}{}_2F_{1}\left(1,\frac{1}{\nu},1+\frac{1}{\nu},-\left(\frac{r}{R}\right)^{\nu}\right)
\end{equation}
where ${}_2F_{1}$ denotes the hypergeometric function. The asymptotic behaviour of the above metric function is given by
\begin{equation}
    f(r)\approx1-8\pi R^2\rho_{0}
\end{equation}
since, for $\nu>0$, $\lim_{r\to \infty}{}_2F_{1}\left(1,\frac{1}{\nu},1+\frac{1}{\nu},-\left(\frac{r}{R}\right)^{\nu}\right)\approx 0$. Thus, the above geometry is not asymptotically Minkowski and has a 
global solid angle deficit for all positive values of $\nu$. 

\noindent The meaning of the solid angle deficit is best understood if we perform the following global coordinate transformation,
\begin{align}\label{4.5}
    \Tilde{r}=\frac{r}{\sqrt{1-8\pi R^2\rho_{0}}}, \hspace{1cm} \Tilde{t}=t\sqrt{1-8\pi R^2\rho_{0}} 
\end{align}
Note that the above transformation is only allowed when $8\pi R^2\rho_{0}<1$. The transformed metric is given as,
\begin{equation}
    ds^2=-\Tilde{f}(\Tilde{r})d\Tilde{t}^2+\frac{d\Tilde{r}^2}{\Tilde{f}(\Tilde{r})}+(1-8\pi R^2\rho_{0})\Tilde{r}^2\left(d\theta^{2}+\sin^{2}{\theta} d\phi^{2}\right)
\end{equation}
where the new metric function $\Tilde{f}(\Tilde{r})$ reaches unity asymptotically (not shown explicitly).
It is now evident that $8\pi R^2\rho_{0}$ has a connection with the solid angle deficit. The surface area of the spherical surface with radius $\Tilde{r}$ is now $4\pi(1-8\pi R^2\rho_{0})\Tilde{r}^2$, which is less than the surface area of the entire sphere. 
In summary, depending on the $\nu$, one can construct different, new spacetime solutions, all of which have a solid angle deficit.
Till now, we have not addressed the regularity of the geometry. 
The previously imposed constraints on the density parameters, aimed at ensuring the regularity of the Ricci scalar and the Ricci contraction
(see Section \ref{II}) in the context of the parameterised density profile Eq.(\ref{2.5}), can be reformulated as the condition $\nu\geq 2$ for the defect subclass. We will calculate the
expressions for the relevant curvature scalars for the specific
example discussed below. We now choose a specific parameter value which will lead to a regular solution characterised by a solid angle deficit.

\subsection{Regular defect spacetime and its geometry}
\begin{figure}[h]
\centering
\includegraphics[width=0.6\textwidth]{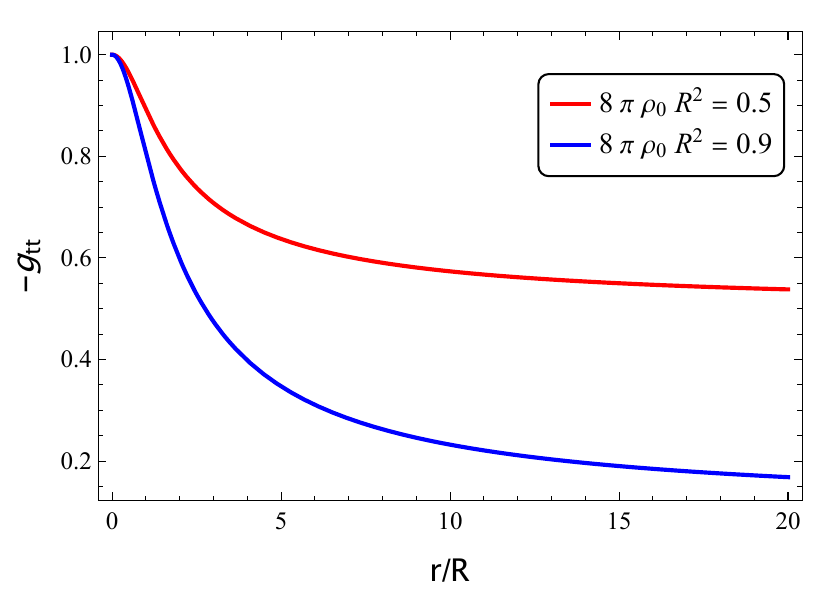}
\caption{Plot of the redshift function of the defect geometry with $r/R$}
\label{fig:mf_rd}
\end{figure}

\noindent Let us consider $\nu=2$, which leads to the following density profile
\begin{equation}
    \rho=\frac{\rho_{0}R^{2}}{r^2+R^2}
\end{equation}
One can identify such a density with the pseudo-isothermal dark matter profile \cite{Isothermal}. Substituting $\nu=2$ in Eq.(\ref{4.4}), we have the following metric function
\begin{equation}\label{4.11}
    f(r)=1-8\pi R^2\rho_{0}+8\pi R^2\rho_{0}\frac{\tan^{-1}\left(\frac{r}{R}\right)}{\frac{r}{R}}
\end{equation}
For small values of $r$, it behaves like de-Sitter space, i.e.
\begin{equation}
    f(r)\approx1-\frac{8\pi\rho_{0}}{3}r^2, \hspace{1cm} r\to 0
\end{equation}
To understand the causal structure of the above spacetime, we examine the zeros of the redshift function ($-g_{tt}$). Since there are no roots of the horizon equation $g_{tt}=0$ (no zeros of the redshift function), a horizon-like structure is absent, as illustrated in Figure \ref{fig:mf_rd}. Moreover, the metric function $-g_{tt}$ and $g_{rr}$ 
do not reduce to unity asymptotically due to the solid angle deficit.
\\
\noindent {\sf{Curvature scalars and geodesic completeness:}}
\noindent To show that $r=0$ is free from a curvature singularity, we demonstrate below the three independent curvature scalars and their limiting values.

\begin{figure}[h]
\centering
\includegraphics[width=0.6\textwidth]{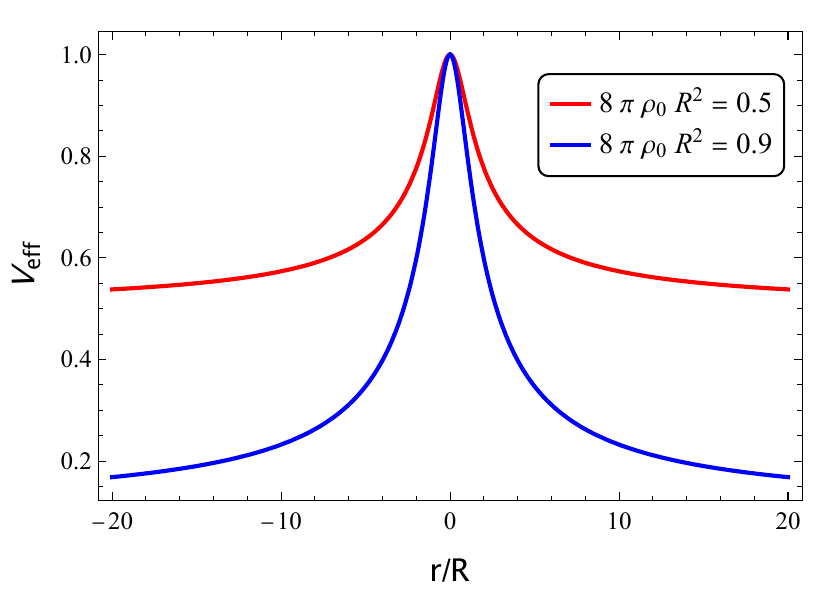}
\caption{Graph illustrating the effective potential for a radial timelike geodesic as a function of $r/R$.}
\label{fig:ep_rd}
\end{figure}

\noindent The Ricci scalar is given as,
\begin{equation}
    g_{\mu\nu}R^{\mu\nu}=\frac{16\pi\rho_{0}R^{2}(r^2+2R^2)}{(r^2+R^2)^2} 
\end{equation}
The Ricci contraction is found to be
\begin{equation}
    R_{\mu\nu}R^{\mu\nu}=\frac{128\pi^2\rho_{0}^2R^2(r^4+2r^2R^2+2R^4)}{(r^2+R^2)^4}
\end{equation}
They have finite values everywhere. The limiting value of the Kretschmann scalar is given as,
\begin{equation}
    \lim_{r\to 0}R_{\mu\nu\lambda\delta}R^{\mu\nu\lambda\delta}=\frac{512\pi^2\rho_{0}^{2}}{3}
\end{equation}
Thus, the defect solution is free from any curvature singularity in the allowed domain of radial coordinate $(0,\infty)$. To show that the geometry is free from a singularity in the extended domain of $r$, i.e.  ($-\infty,\infty$), we have to demonstrate its geodesic completeness.
Similar to the previous example, the effective potential of radial timelike geodesics shows a smooth and continuous behaviour across the extended radial coordinate, as illustrated in Figure \ref{fig:ep_rd}. Therefore, the spacetime is geodesically complete. 
In summary, the above metric characterises a regular compact object without a horizon and possesses a solid angle deficit.

\noindent {\sf{Embedding diagram:}} To visualise the shape of the regular defect spacetime, we embed its two-dimensional sections in an Euclidean background. Specifically, we consider a spatial 2D slice of the spacetime by fixing $t=\text{constant}$ and $\theta=\pi/2$.
This slice is then embedded in three-dimensional Euclidean cylindrical coordinates. The 2D spatial section of the above metric takes the following form,
\begin{equation}\label{4.13}
    ds^2=\frac{d\Tilde{r}^2}{\Tilde{f}(\Tilde{r})}+\Tilde{r}^2d\Tilde{\phi}^2
\end{equation}
where $\Tilde{f}(\Tilde{r})=\frac{1}{1-8\pi R^2\rho_{0}}\left(1-8\pi R^2\rho_{0}+8\pi R^3\rho_{0}\frac{\tan^{-1}\left(\frac{\Tilde{r}\sqrt{1-8\pi R^\rho_{0}}}{R}\right)}{\Tilde{r}\sqrt{1-8\pi R^2\rho_{0}}}\right)$ and $\Tilde{\phi}=\phi\sqrt{1-8\pi R^2\rho_{0}}$. Note that the above metric is written in the transformed coordinate mentioned earlier in Eq.(\ref{4.5}).
The line element in cylindrical coordinate in 3D flat space is,
\begin{equation}\label{4.14}
    ds^2=d{\Tilde r}^2+ {\Tilde r}^2d \Tilde\phi^2+dz^2
\end{equation}
\begin{figure}[h]
\centering
\includegraphics[width=0.6\textwidth]{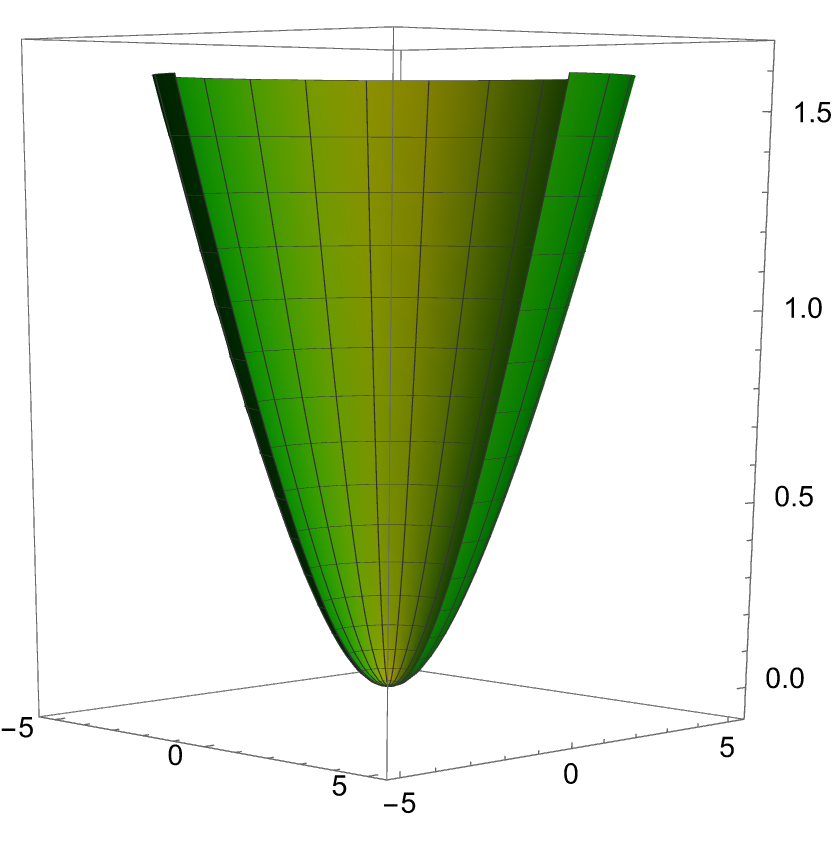}
\caption{Embedding diagram of the defect geometry for $8\pi R^2\rho_0=0.5$.}
\label{fig:Em_df}
\end{figure}
Defining  a profile function $z(\Tilde r)$ and comparing Eq. (\ref{4.13}) 
with the two dimensional $z= z(\Tilde r)$ section of Eq. (\ref{4.14}), we obtain the profile function $z(\Tilde{r})$. Rotating the profile function
over the range  ($0, 2\pi \sqrt{1-8\pi R^2 \rho_0}$) of $\Tilde \phi$ (note the range
is not $(0,2\pi)$) we obtain the shape of the embedded slice 
(for a choice of $8\pi R^2\rho_0=0.5$) as shown in Figure \ref{fig:Em_df}. In the full four dimensional sense, the geometry has a {\em solid angle deficit}. However, the embedding diagram which 
represents a 2d slice of the original metric captures an angular deficit 
only which still demonstrates the reduced range of the azimuthal coordinate $(\Tilde{\phi})$. If we had chosen to work with the 
original coordinates $(r,\phi)$ we would have obtained a deformed conical shape with a flattened vertex and without the gap which appears in Fig. \ref{fig:Em_df} due the
reduced range of $\Tilde \phi$.

\noindent One may also understand this character of the
geometry by recalling the simple example of a two dimensional line element given as $ds^2 = \frac{dr^2}{1-\alpha} + r^2 d\phi^2$ with $\alpha<1$. Embedding this geometry in a 3D cylindrical background we obtain a profile function
$z(r) \propto r$ which represents a cone. In contrast, if we scale the coordinates $r$ and $\phi$ such that $r= \sqrt{1-\alpha} \, r'$ and $\phi'=
\sqrt{1-\alpha} \, \phi$, the metric goes over to $ds^2 = dr'^2 + r'^2 d\phi'^2$ with the range of $\phi'$ reduced. Thus, in $r'\phi'$, we just get flat space with a gap, which is the conical deficit. In our specific example both the $r\phi$ and $r'\phi'$ representations appear curved 
in shape (due to the nature of the metric) with the latter directly demonstrating the presence of an angular deficit (in reality, a
solid angle deficit in full 3D). 
This, in brief, is the central notion of a spacetime with a defect, 
illustrated and discussed extensively in the past in studies on global monopoles or cosmic strings.

\noindent In addition, the regular nature of the geometry at $r=0$ is also evident in the embedding diagram, and, as we shall see below, the above structure may be used to model the interior of a compact stellar object.

\begin{figure}[h]
\centering
\includegraphics[width=0.6\textwidth]{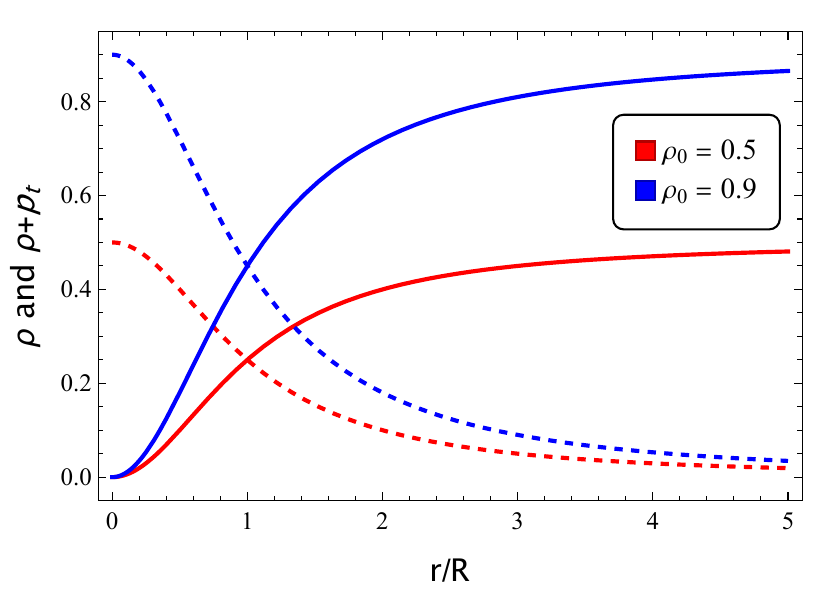}
\caption{Graph of $\rho$ and $\rho + p_t$ as a function of $r/R$ for various values of $\rho_{0}$.  The dashed and solid lines denote $\rho$ and $\rho + p_t$, respectively.  Lines of the same colour possess same $\rho_{0}$ values.}
\label{fig:NEC_df}
\end{figure}

\begin{figure}[h]
\centering
 \includegraphics[width=0.6\textwidth]{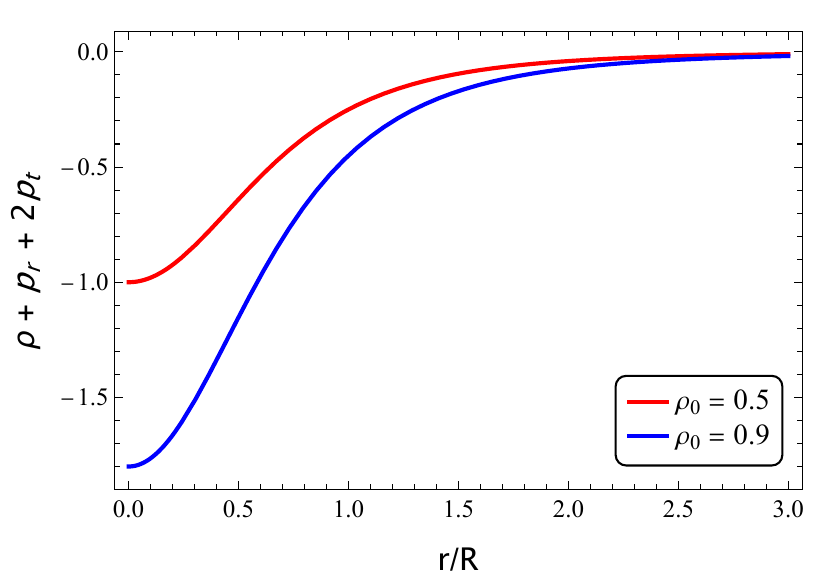}
\caption{Graph of the L. H. S. of the SEC with $r/R$, for various values of $\rho_{0}$.}
\label{fig:SEC_df}
\end{figure}

\noindent {\sf{Energy conditions:}}
\noindent Let us now move on to demonstrating the status of the different energy conditions (assuming Einsteinian GR) for the matter required to support such a geometry.
The diagonal elements of the energy-momentum tensor (in the frame basis) are the following,
\begin{align}
    \rho=-p_r=\frac{\rho_{0}R^{2}}{r^2+R^2}, \hspace{1cm} p_t=-\frac{\rho_{0}R^4}{(r^2+R^2)^2}
\end{align}
where $\rho$ is assumed previously. It is evident from the above equations that
\begin{align}
    \rho>0, \hspace{1cm} \rho+p_r=0, \hspace{1cm} \rho +p_t=\frac{\rho_{0}R^2r^2}{(r^2+R^2)}>0
\end{align}
\noindent Thus, NEC and WEC hold over the entire domain of the radial coordinate, which is also evident from Figure \ref{fig:NEC_df}.
However, the SEC inequality is violated in the full domain of $r$, as shown in
Eq.(\ref{4.17}) below and Figure \ref{fig:SEC_df}.
\begin{equation}\label{4.17}
    \rho+p_r+2p_t=-\frac{2\rho_{0}R^4}{(r^2+R^2)^2}<0
\end{equation}

\subsection{Lagrangian model for the required matter}
\noindent In the GR coupled to matter scenario, we may use a fluid of strings as the possible matter model which can support such a regular defect geometry. The idea of a fluid of strings is an extension of the older idea
of a cloud of strings where pressure is incorporated. The model of a cloud of strings is based on a surface bivector $\Sigma^{\mu\nu}$ that spans the 2D timelike worldsheet of strings \cite{Letelier1}. This is given as,
\begin{equation}
    \Sigma^{\mu\nu}=\epsilon^{AB}\frac{\partial x^{\mu}}{\partial\zeta^{A}}\frac{\partial x^{\nu}}{\partial\zeta^{B}}
\end{equation}
where $\epsilon^{AB}$ is the 2D Levi-Civita symbol. The worldsheet coordinates are $\zeta^{0}$ and $\zeta^1$, which are timelike and spacelike, respectively.
The induced metric on the worldsheet is written as
\begin{equation}
    h_{AB}=g_{\mu\nu}\frac{\partial x^\mu}{\partial \zeta^A}\frac{\partial x^\nu}{\partial \zeta^B}
\end{equation}
The energy-momentum tensor of the cloud of strings, as proposed by Letelier \cite{Letelier1}, is characterised by an energy density $\rho$ and is
generally given as,
\begin{equation}
    T^{\mu\nu}=\rho \sqrt{-h}\frac{\Sigma^{\mu\lambda}\Sigma^{\nu}_{\lambda}}{(-h)}
\end{equation}
where $h$ is the determinant of the induced metric. The generalisation to a fluid of strings is done by including pressure \cite{Letelier2}. The energy-momentum tensor for a perfect fluid of strings is
\begin{equation}
    T^{\mu\nu}=\left(p+\rho\sqrt{-h}\right)\frac{\Sigma^{\mu\lambda}\Sigma^{\nu}{}_{\lambda}}{(-h)}+p g^{\mu\nu}
\end{equation}
where $p$ and $\rho$ are the pressure and density of the fluid of strings, respectively. Now, our goal is to associate the fluid of strings as a source for the regular defect geometry. As our regular defect geometry described in Eq.(\ref{4.11}) is static, spherically symmetric and in the Schwarzschild gauge, the only nonvanishing components of $\Sigma_{\mu\lambda}$ are $\Sigma_{tr}$ and $\Sigma_{\theta\phi}$ \cite{Soleng}. As a result, the diagonal elements of the above energy-momentum tensor become
\begin{equation}
    T^{\mu}{}_{\nu}=\left[-\rho(r),-\rho(r),p,p\right]
\end{equation}
For the regular defect geometry in Eq.(\ref{4.11}), we have the following components of the energy-momentum tensor (radial pressure is negative and equal in magnitude to the density),
\begin{align}
    \rho=\frac{\rho_{0}R^{2}}{r^2+R^2}, & \hspace{1cm} p=-\frac{\rho_{0}R^4}{(r^2+R^2)^2}
\end{align}
which can be mapped onto the energy-momentum tensor for a fluid of strings.
Therefore, one may relate the matter energy-momentum tensor required to support the defect geometry with the fluid of strings. Note that the equation of state for the fluid of strings is,
\begin{equation}
    p=-\frac{1}{\rho_{0}}\rho^2
\end{equation}
which is a polytropic equation of state ($p\propto\rho^2$).
In the literature,  different equations of state for a fluid of strings have been studied and the corresponding spacetime solutions constructed. In some cases, pressure and density are related by a constant factor \cite{Soleng}. Works based on an arbitrary function of $r$ that connects pressure and density can also be found \cite{Santos}. 
In our study, we require a fluid of strings which satisfies the polytropic equation of state. 
It is interesting to note that asymptotically ($r>>R$), the density and pressure become
\begin{equation}
    \rho\approx\frac{\rho_{0}R^2}{r^2},\hspace{0.5cm} \text{and} \hspace{0.5cm}p\to 0
\end{equation}
where we consider the terms up to $O(\frac{1}{r^2})$. As a result, the metric function becomes
\begin{equation}
    f(r)\approx1-8\pi R^2\rho_{0}
\end{equation}
which is the solution for the cloud of strings. Thus, one can summarise that the regular defect geometry in Eq.(\ref{4.11}) can be obtained in the presence of a polytropic fluid of strings (in GR). Asymptotically, the geometry represents a flat spacetime surrounded by a cloud of strings, which may be a reason behind the appearance of the solid angle deficit.

\noindent We would now like to use the spacetime introduced here in constructing a model of a stable star.


\section{Model of a stable star (gravastar)}\label{VI}
\noindent In GR, stellar models are constructed
by considering a spherical region (of radius $R$ say) with the matter inside obeying a certain equation of state. Thus, via Einstein's equations (or the Tolman--Oppenheimer--Volkoff equation) one can obtain the metric functions.
The simplest example is that of a constant density star which is 
discussed in many standard texts on GR (eg. see \cite{Wald}). The boundary of this compact region is then matched with an exterior vacuum
which, for the non-rotating case, is taken as Schwarzschild spacetime.
Obviously, the metrics and their derivatives  (extrinsic curvature components) must match at
the boundary. These matching conditions (similar to conditions on
electric and magnetic fields at the interface between two media, in electrodynamics), in the general scenario,
are the Sen-Israel-Darmois junction conditions \cite{sen, Israel, Darmois}. In many situations, a direct matching is not possible and one ends up with a thin shell at the boundary which carries a density and a pressure with a specific equation of state. At the next level,
one needs to study the stability of this construction by making the
boundary dynamic and studying its evolution.

\noindent The regular defect spacetime in Eq.(\ref{4.11}) possesses the following key features: (a) it behaves like de-Sitter space at the centre, (b) its embedding diagram (Fig.\ref{fig:Em_df}) illustrates that its geometry is like the interior of a star and (c) a polytropic fluid of strings can model the required matter. These properties motivate us to construct a stable star using the regular defect geometry as its interior. For this purpose, we consider the Visser--Wiltshire dynamically stable thin shell model \cite{Visser1}, which is based on the
junction condition formalism. The interior defect geometry joins with the outer Schwarzschild metric via a thin shell. The thin shell behaves like the joining surface between the two spacetimes \cite{Visser2, Visser3}. This three-layer model may be described as follows:

$\bullet$ An outer Schwarzschild geometry representing vacuum.

$\bullet$ A thin shell with specific surface density and surface tension.

$\bullet$ The interior regular defect geometry.

\noindent Thus, two different spacetimes are joined at a surface ($\Sigma$) to form a single spacetime. 

\begin{figure}[h]
\centering
 \includegraphics[width=0.43\textwidth]{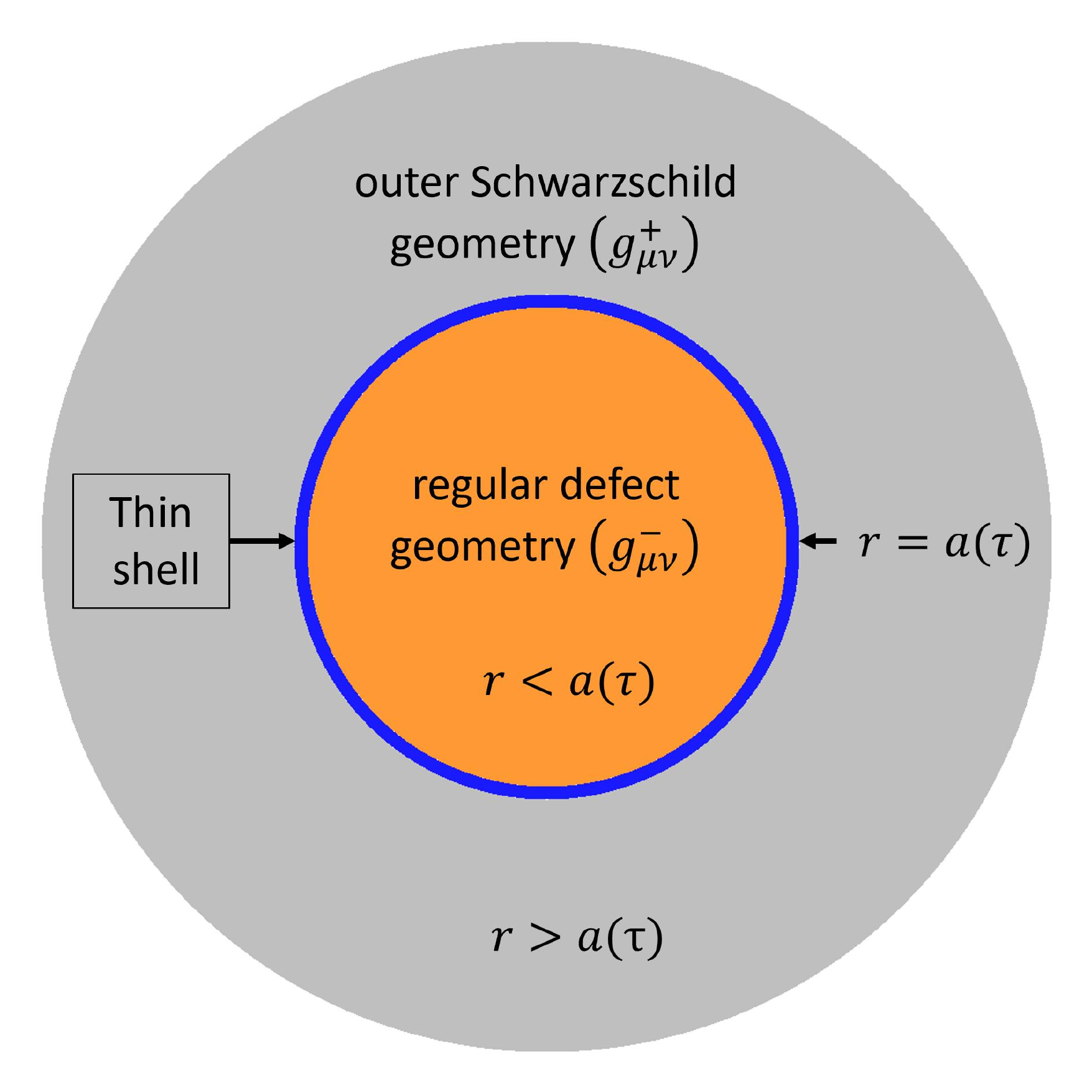}
\caption{Qualitative picture of the thin shell star}
\label{fig:Star}
\end{figure}

\noindent We begin our construction of the stable star assuming a general interior spacetime ($g_{\mu\nu}^{-}$) and a general exterior solution ($g^{+}_{\mu\nu}$). The signs $+$ and $-$ indicate the respective outer and inner solutions.  In Figure \ref{fig:Star}, we have given a 
qualitative sketch of our thin shell star (gravastar) model. 
To avoid the horizon in the exterior, we consider the junction radius to be larger than the Schwarzschild radius.

\noindent The line element of the interior metric $(g_{\mu\nu}^{-})$ is written as
\begin{equation}
    ds^{2}=-f_{-}(r)dt^{2}+\frac{dr^2}{f_{-}(r)}+r^2(d\theta^2+\sin^2{\theta} d\phi^2)
\end{equation}
where $f_{-}(r)=1-\frac{2m_{-}(r)}{r}$. The exterior metric $(g_{\mu\nu}^{+})$ is
\begin{equation}
    ds^{2}=-f_{+}(r)dt^{2}+\frac{dr^2}{f_{+}(r)}+r^2(d\theta^2+\sin^2{\theta} d\phi^2)
\end{equation}
where $f_{+}(r)=1-\frac{2m_{+}(r)}{r}$

\noindent We assume that the above two geometries are connected along a dynamical timelike hypersurface ($\Sigma$) at $r=a(t)$, with a spacelike normal.

\noindent Below, we first review the junction
condition formalism in the context of our work. For a useful
pedagogical review of the 
junction condition formalism see \cite{poisson}. In the next subsection, we apply the general formalism to our case and evaluate the required surface density and surface tension of the thin shell by 
using the corresponding metric functions in the two regions.

\subsection{Junction conditions and formalism}
\noindent The first junction condition states that the metric must be continuous across the boundary. Hence, the induced metric on the junction surface observed from interior and exterior cannot have a discontinuity, i.e.
\begin{equation}
    g^{+}_{ij}(\xi)=g^{-}_{ij}(\xi)=g_{ij}(\xi)
\end{equation}
Here, $i=1,2,3$ represents the index of basis vectors tangent to the hypersurface, and $\mu=1,2,3,4$ is the spacetime index. The three basis vectors $e_{(i)}=\partial/\partial\xi^{i}$ tangent to $\Sigma$ have the following components: $e_{(i)}^{\mu}|_{\pm}=\partial x_{\pm}^{\mu}/\partial\xi^{i}$, which are used to construct the induced metric on the junction surface given as,
\begin{equation}
    g_{ij}=g_{\mu\nu}e_{(i)}^{\mu}e_{(j)}^{\nu}|_{\pm}
\end{equation}
The natural choice of the coordinates of the junction surface is $\xi^{i}=(\tau,\theta,\phi)$, where $\tau$ is the proper time along the hypersurface $\Sigma$. 
Considering a point with fixed $\theta$ and $\phi$ on the hypersurface, the proper time defined from the inner and outer metrics give
 \begin{equation}\label{5.5}
 -d\tau^2=-f_{\pm}(a)dt_{\pm}^2+\frac{1}{f_{\pm}(a)}\left(\frac{da}{dt_{\pm}}\right)^2dt_{\pm}^2
 \end{equation}
Therefore, the three basis vectors tangent to the hypersurface are:
\begin{equation}
    \begin{aligned}
    e^{\mu}_{(\tau)}=(\dot{t_{\pm}},\dot{a},0,0), \hspace{1cm}
     e^{\mu}_{(\theta)}=(0,0,1,0), \hspace{1cm}
      e^{\mu}_{(\phi)}=(0,0,0,1)
    \end{aligned}
\end{equation}
where an overdot represents derivative with respect to proper time ($\tau$). The induced metrics on the hypersurface computed from the exterior and interior metrics are the following:
\begin{align}
    g_{ij}^{\pm}d\xi^{i}d\xi^{j}=\left(-f_{\pm}(a)\dot{t_{\pm}}^2 +\frac{\dot{a}^2}{f_{\pm}(a)}\right)d\tau^2+a^2(\tau)(d\theta^2+\sin^2{\theta} d\phi^2)
\end{align}
From the first junction condition, we have
\begin{equation}
    -f_{-}(a)\dot{t_{-}}^2+\frac{\dot{a}^2}{f_{-}(a)}=-f_{+}(a)\dot{t_{+}}^2+\frac{\dot{a}^2}{f_{+}(a)}
\end{equation}
which holds trivially from the definition of proper time in Eq.(\ref{5.5}). Moreover, on the hypersurface we have: $-f_{\pm}(a)\dot{t_{\pm}}^2 +\frac{\dot{a}^2}{f_{\pm}(a)}=-1$. Thus, the induced metric becomes
\begin{equation}
    g_{ij}d\xi^{i}d\xi^{j}=-d\tau^2+a^2(\tau)(d\theta^2+\sin^2{\theta} d\phi^2)
\end{equation}
Now the position of the junction surface is given by $x^{\mu}(\tau,\theta,\phi)=(t(\tau), a(\tau),\theta,\phi)$ and the 4-velocity as measured from the two sides of the junction are,
\begin{equation}
    U_{\pm}^{\mu}=\left(\frac{\sqrt{f_{\pm}(a)+\dot{a}^2}}{f_{\pm}(a)},\dot{a},0,0\right)
\end{equation}
The unit normal to the junction surface can be obtained as,
\begin{equation}
    n^{\pm}_{\mu}=\left(-\dot{a},\frac{\sqrt{f_{\pm}(a)+\dot{a}^2}}{f_{\pm}(a)},0,0\right)
\end{equation}
which obeys $U^{\mu}n_{\mu}=0$ and $n^{\mu}n_{\mu}=1$. 
One may note some key differences in the directions of the normal vectors as one approaches the junction in a thin-shell wormhole or in a thin-shell gravastar/star. In a thin-shell gravastar/star, the normal vectors,
as one approaches the junction from either geometry are directed towards the junction surface \cite{Visser5}. However, the thin shell wormhole consists of two 
identical geometries glued at the junction surface. As a result, the two normal vectors are along opposite directions w.r.t. the junction. This leads to an extra negative sign in the normal vector w.r.t. one of the geometries \cite{Visser4}. 

\noindent The extrinsic curvature ( i.e. the curvature of the embedded surface, encoded in the gradient of the normal to the surface) is defined as $K_{ij}=\nabla_{\nu}n_{\mu}e^{\mu}_{(i)}e^{\nu}_{(j)}$, or
\begin{equation}
    K^{\pm}_{ij}=-n_{\mu}^{\pm}\left(\frac{\partial^2 x^{\mu}}{\partial\xi^{i}\partial\xi^{j}}+\Gamma^{\mu\pm}_{\alpha\beta}\frac{\partial x^{\alpha}}{\partial \xi^{i}}\frac{\partial x^{\beta}}{\partial \xi^{j}}\right)
\end{equation}
The diagonal components of the extrinsic curvature (using the above expressions for the basis and the normal) are given as follows \cite{Lobo, Bhattacharjee}:
\begin{equation}
    \begin{aligned}
        K^{\tau\pm}_{\tau}=\frac{\ddot{a}+f_{\pm}^{\prime}(a)/2}{\sqrt{f_{\pm}(a)+\dot{a}^2}}\\
        K^{\theta\pm}_{\theta}=K^{\phi\pm}_{\phi}=\frac{1}{a}\sqrt{f_{\pm}(a)+\dot{a}^{2}}
    \end{aligned}
\end{equation}
The second junction condition states that the jump in the extrinsic curvature is directly
proportional to the surface energy-momentum tensor $S_{ij}$ at the shell \cite{Lanczos}. Explicitly, we have
\begin{equation}
    [[K_{ij}-K g_{ij}]]=-8\pi S_{ij}
\end{equation}
where the notation $[[O]] = O^+- O^{-}$ represents the jump of $O$ at the junction and $K$ is the trace of $K_{ij}$. The diagonal elements of the surface energy-momentum tensor can be written as $S^{i}_{j}=diag(-\sigma,-v,-v)$. Thus, we have
\begin{equation}\label{5.15}
    \sigma=-\frac{1}{4\pi}(K^{\theta +}_{\theta}-K^{\theta -}_{\theta})
\end{equation}
and,
\begin{equation}\label{5.16}
    v=-\frac{1}{8\pi}(K^{\tau +}_{\tau}+K^{\theta +}_{\theta}-K^{\tau -}_{\tau}-K^{\theta -}_{\theta})
\end{equation}
where $\sigma$ and $v$ represent the surface energy and surface tension, respectively. If the surface stress-energy term vanishes, then the junction surface can be treated as a boundary. The finite stress energy makes the junction a thin shell. Now, using Eq.(\ref{5.15}) and (\ref{5.16}), a relation between $\sigma$ and $v$ can be found, 
\begin{equation}\label{5.17}
    \frac{d}{d\tau}(\sigma a^2)=v\frac{d}{d\tau}(a^2)
\end{equation}
which may be identified as the conservation equation of energy or the continuity equation of the fluid of matter within the thin shell.

\noindent Thus, if we have the equation of state of the shell $v=v(\sigma)$, one can obtain $\sigma(a)$ as a function of $a$ from the conservation equation. As a result, the full dynamical behaviour of $a$ is encoded in a single equation (\ref{5.15}).

\subsection{Master equation}
\noindent To obtain the dynamical stability of the thin shell at $r=a(\tau)$, we may proceed in the following two ways \cite{Visser1}:

\noindent {\sf{First procedure:}}
\noindent The dynamical equation for $\sigma$ in Eq.(\ref{5.15}) can be rewritten as,
\begin{equation}\label{5.18}
    \sqrt{1-2m_{+}(a)/a+\dot{a}^2}=\sqrt{1-2m_{-}(a)/a+\dot{a}^2}-4\pi a\sigma(a)
\end{equation}
Note that we have a positive $\sigma(a)$ for $m_{+}(a)>m_{-}(a)$. A star can be constructed of null surface energy density for specific parameter values when $m_{+}(a)=m_{-}(a)$. In contrast, a thin shell wormhole always requires a finite surface energy density due to the sign flip of the normal vectors discussed earlier.

\noindent The above master equation may be re-expressed using a potential as,
\begin{equation}
    \frac{\dot{a}^2}{2}+V(a)=0
\end{equation}
where, 
\begin{equation}
    V(a)=\frac{1}{2}\left(1+\frac{4m_{+}(a)m_{-}(a)}{m_{s}^2(a)}-\left(\frac{m_{s}(a)}{2a}+\frac{m_{+}(a)+m_{-}(a)}{m_{s}(a)}\right)^2\right)
\end{equation}
Here, $m_{s}(a)=4\pi\sigma(a)a^2$. 
There will exist a strictly stable solution for the shell if and only if there is some $m_{s}$ and some $a_{0}$ such that,
\begin{equation}
    V(a_{0})=0, \hspace{1cm} V^{\prime}(a_{0})=0, \hspace{1cm} V^{\prime\prime}(a_0)>0
\end{equation}
Thus, given an equation of state $v=v(\sigma)$ and the conservation equation, we can find out $\sigma=\sigma(a)$. Next, we can find the stable point $(a_0)$ for the junction surface by analysing the stability of the $V(a)$. 

\noindent {\sf{Alternate procedure:}}
\noindent An alternate way involves writing $\sigma(a)$ in terms of $V(a)$. We assume that we know the potential $V(a)$ having a stable point $(a_0)$. Therefore Eq.(\ref{5.18}) becomes,
\begin{equation}\label{5.22}
    \sigma(a)=-\frac{1}{4\pi a}\left(\sqrt{1-2V(a)-\frac{2m_{+}(a)}{a}}-\sqrt{1-2V(a)-\frac{2m_{-}(a)}{a}}\right)
\end{equation}
The differences between Eqs. (\ref{5.18}) and (\ref{5.22}) are the following: (a) Eq.(\ref{5.18}) provides the surface energy on a shell with velocity $\dot{a}$ at radius $a$; (b) Eq.(\ref{5.22}) demonstrates what the surface energy has to be, so that it is compatible with the chosen potential $V(a)$. Therefore, Eq.(\ref{5.22}) is more powerful 
in a dynamical analysis.

\noindent From the conservation equation (\ref{5.17}), one can obtain $v(a)$ as follows,
\begin{equation}
    \begin{aligned}
        v(a)=-\frac{1}{8\pi a}\left(\frac{1-2V(a)-m_{+}(a)/a-a V^{\prime}(a)-m_{+}^{\prime}(a)}{\sqrt{1-2V(a)-\frac{2m_{+}(a)}{a}}} \right. \\ \left.-\frac{1-2V(a)-m_{-}(a)/a-a V^{\prime}(a)-m_{-}^{\prime}(a)}{\sqrt{1-2V(a)-\frac{2m_{-}(a)}{a}}}\right)
    \end{aligned} 
\end{equation}  
Here, we choose a $V(a)$ having a stable point. Then, one can find out the $\sigma(a)$, $v(a)$ and the corresponding equation of state.

\subsection{Our system}
\begin{figure}[!htbp]
\centering
\subfigure[\hspace{0.1cm}$8\pi R^2\rho_{0}=0.5$]{\includegraphics[width=0.45\textwidth]{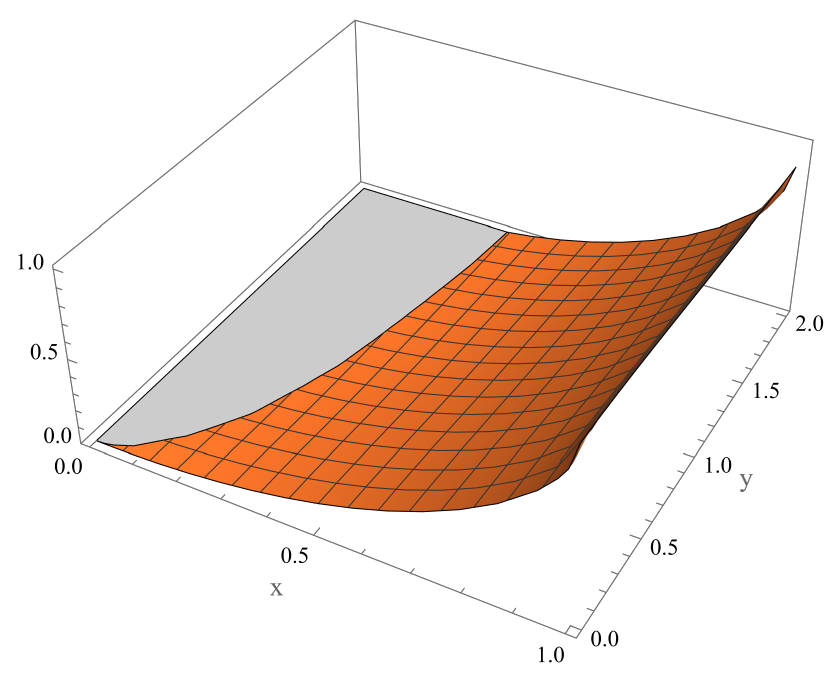}\label{subfig:Surfaceenergy1}}
\subfigure[\hspace{0.1cm}$8\pi R^2\rho_{0}=0.9$]{\includegraphics[width=0.45\textwidth]{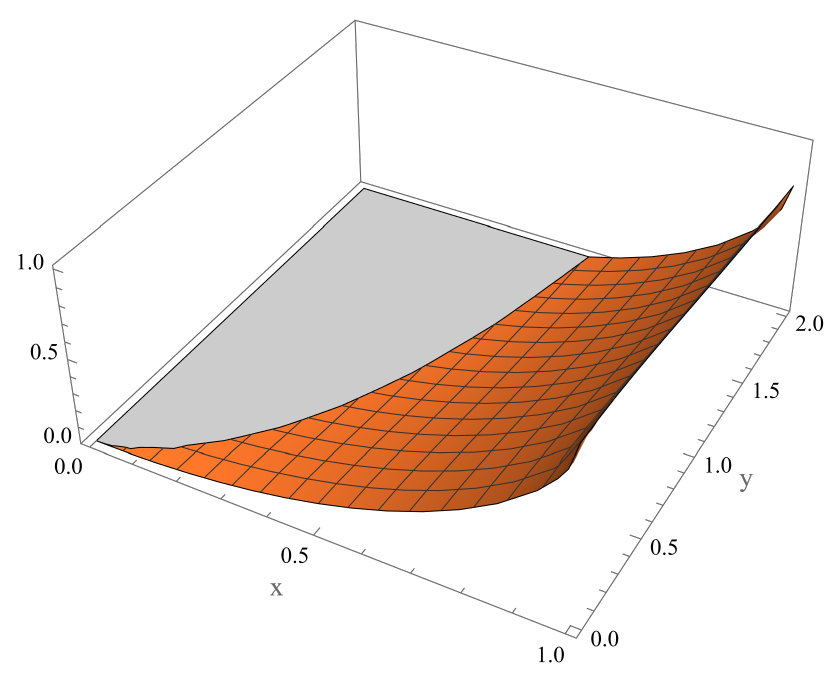}\label{subfig:Surfaceenergy2}}
\caption{A qualitative representation of $\mu(a_0)$ within the parameter space $(x,y)$ for $8\pi R^2\rho_{0}=0.5$ (left) and $8\pi R^2\rho_{0}=0.9$ (right).  The parameter $x$ ranges from $0$ to $1$, while $y$ may assume any positive value.  The graphs illustrate the areas where $\mu(a_0)$ is positive (shown in orange), whereas places with negative $\mu(a_0)$ are omitted (depicted in grey).}
\label{fig:surfaceenergy}
\end{figure}

\begin{figure}[!htbp]
\centering
\subfigure[\hspace{0.1cm}$8\pi R^2\rho_{0}=0.5$]{\includegraphics[width=0.45\textwidth]{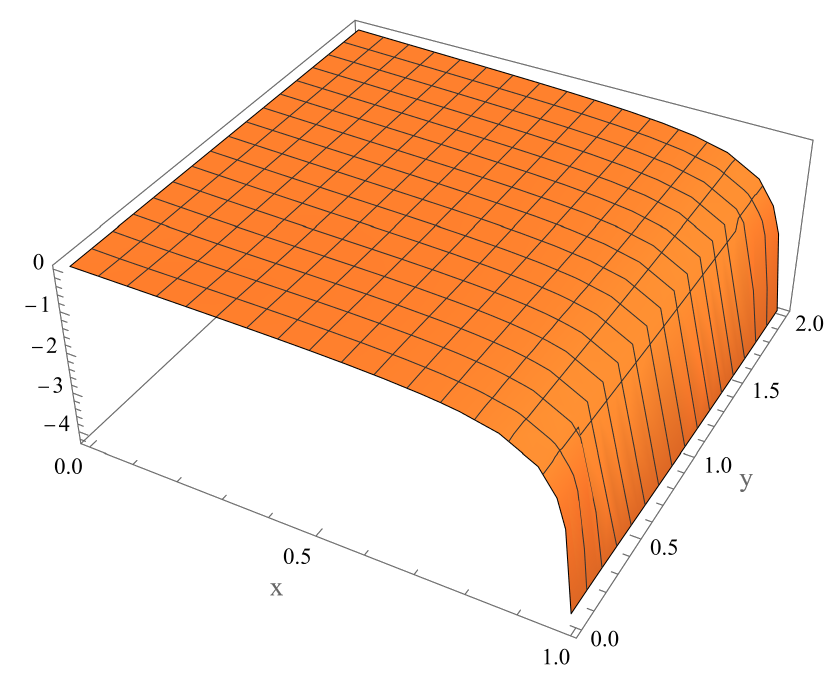}\label{subfig:Surfacetension1}}
\subfigure[\hspace{0.1cm}$8\pi R^2\rho_{0}=0.9$]{\includegraphics[width=0.45\textwidth]{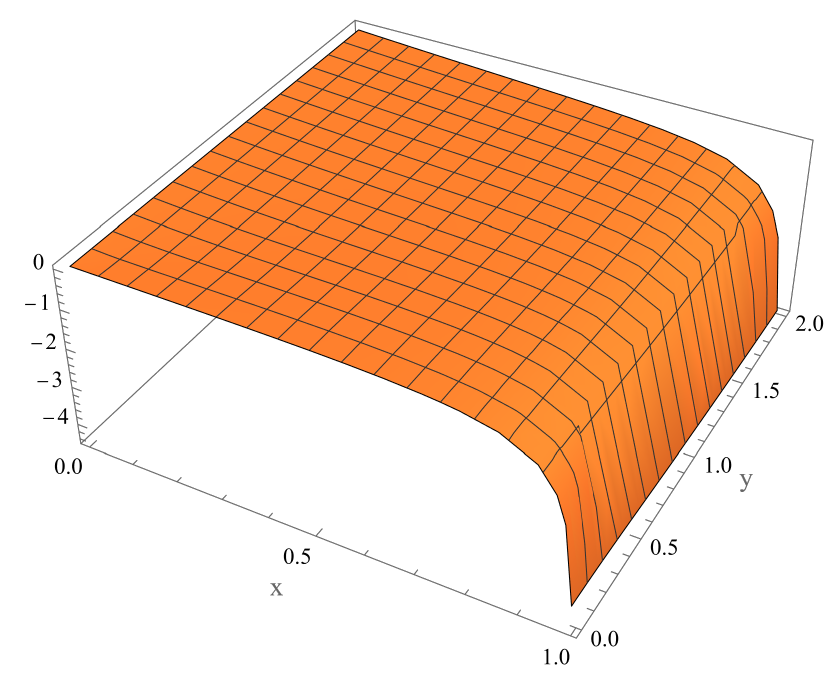}\label{subfig:Surfacetension2}}
\caption{Variation of $\Pi(a_0)$ within the parameter space $(x,y)$ for $8\pi R^2\rho_{0}=0.5$ (left) and $8\pi R^2\rho_{0}=0.9$ (right).  The parameter $x$ ranges from $0$ to $1$, while $y$ may assume any positive value.}
\label{fig:surfacetension}
\end{figure}

\noindent We will now calculate the surface energy density and surface tension at the shell by considering the interior as the regular defect geometry and the exterior as a Schwarzschild geometry. From these quantities, the equation of state at the shell will also be found. For this purpose, we choose to work with the alternate procedure discussed above. The interior and exterior mass functions are as follows:
\begin{equation}
    m_{+}(a)=M, \hspace{1cm} m_{-}(a)=4\pi R^2\rho_{0}a-4\pi R^3\rho_{0}\tan^{-1}(a/R)
\end{equation}
Therefore, the surface energy density is obtained as
\begin{equation}\label{5.25}
    \sigma(a)=\frac{1}{4\pi a}\left(\sqrt{1-2V(a)-8\pi R^2\rho_{0}+8\pi R^3\rho_{0}\frac{\tan^{-1}(a/R)}{a}}-\sqrt{1-2V(a)-\frac{2M}{a}}\right)
\end{equation}
and the surface tension is
\begin{equation}\label{5.26}
    \begin{aligned}
        v(a)=\frac{1}{8\pi a}\left(\frac{1-2V(a)-4\pi R^2\rho_{0}+4\pi R^3\rho_{0}\frac{\tan^{-1}(a/R)}{a}-aV^{\prime}(a)-4\pi \rho_{0}R^2\frac{a^2}{(a^2+R^2)}}{\sqrt{1-2V(a)-8\pi R^2\rho_{0}+8\pi R^3\rho_{0}\frac{\tan^{-1}(a/R)}{a}}}\right.\\ \left.-\frac{1-2V(a)-M/a-aV^{\prime}(a)}{\sqrt{1-2V(a)-\frac{2M}{a}}}\right)
    \end{aligned}
\end{equation}
\noindent The surface energy density and surface tension described in Eq. (\ref{5.25}) and (\ref{5.26}) determine the equation of state of the matter in the shell in terms of a freely specified $V(a)$.
Now, if we choose $V(a)=0$, any constant value of $a$ ($a>2M$) represents a stable point $(a_0)$, which also represents a static star model. Considering suitable dimensionless parameters, the surface energy density and surface tension of the static star may be expressed as follows,
\begin{equation}\label{5.27}
    \mu(a_0)=x\left(\sqrt{1-8\pi R^2\rho_{0}+8\pi R^2\rho_{0}\frac{\tan^{-1}(y/x)}{y/x}}-\sqrt{1-x}\right)
\end{equation}
\begin{equation}\label{5.28}
    \begin{aligned}
        \Pi(a_0)=x\left(\frac{1-4\pi R^2\rho_{0}+4\pi R^2\rho_{0}\frac{\tan^{-1}(y/x)}{y/x}-\frac{4\pi R^2\rho_{0}}{1+(x/y)^2}}{\sqrt{1-8\pi R^2\rho_{0}+8\pi R^2\rho_{0}\frac{\tan^{-1}(y/x)}{y/x}}}-\frac{1-x/2}{\sqrt{1-x}}\right)
    \end{aligned}
\end{equation}
where we define $\mu(a_0)=8\pi M\sigma(a_0)$, $x=2M/a_0$, $y=2M/R$ and $\Pi(a_0)=16\pi M v(a_0)$. Note that the above quantities have real value when $x<1$ and $8\pi R^2\rho_{0}<1$. Moreover, $\mu(a_0)$ is positive when
\begin{equation}
    8\pi R^2\rho_{0}\left(1-\frac{\tan^{-1}(y/x)}{y/x}\right)<x
\end{equation}
The above expression represents a transcendental inequality between $y$ and $x$. However, for all possible values of $y$, we have positive $\mu(a_0)$ when $8\pi \rho_{0}R^2<x$. In Figure \ref{fig:surfaceenergy}, we qualitatively demonstrate the parametrised surface energy density $(\mu)$ for different $8\pi R^2\rho_{0}$. The orange region in the parameter space $(x,y)$ represents the positive parametrised surface energy density $(\mu)$ and the grey region corresponds to values where it is negative
and therefore, excluded. Thus, the boundary between the orange and grey regions represents the specific $(x,y)$ where $\sigma(a_0)=0$.
Figure \ref{fig:surfacetension} represents the variation of the parametrised surface tension $(\Pi)$ on the thin shell with $(x,y)$. It is found that $\Pi(a_0)$ is always negative in the full parameter space $(x,y)$ for all the allowed chosen values of $8\pi R^2\rho_{0}$.
Moreover, it reaches $-\infty$ when $x=1$ or $a_0=2M$.

\begin{figure}[!htbp]
\centering
\subfigure[\hspace{0.1cm}$8\pi R^2\rho_{0}=0.5$]{\includegraphics[width=0.45\textwidth]{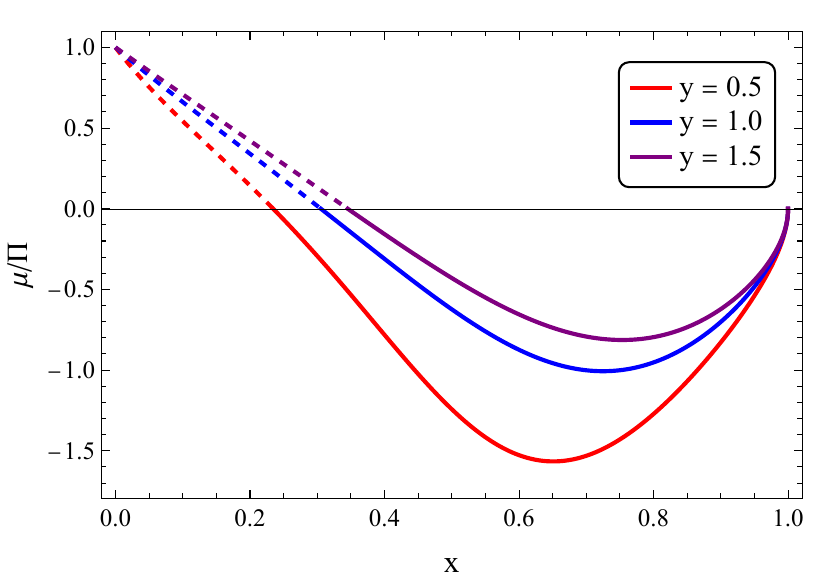}\label{subfig:Eos1}}
\subfigure[\hspace{0.1cm}$8\pi R^2\rho_{0}=0.9$]{\includegraphics[width=0.45\textwidth]{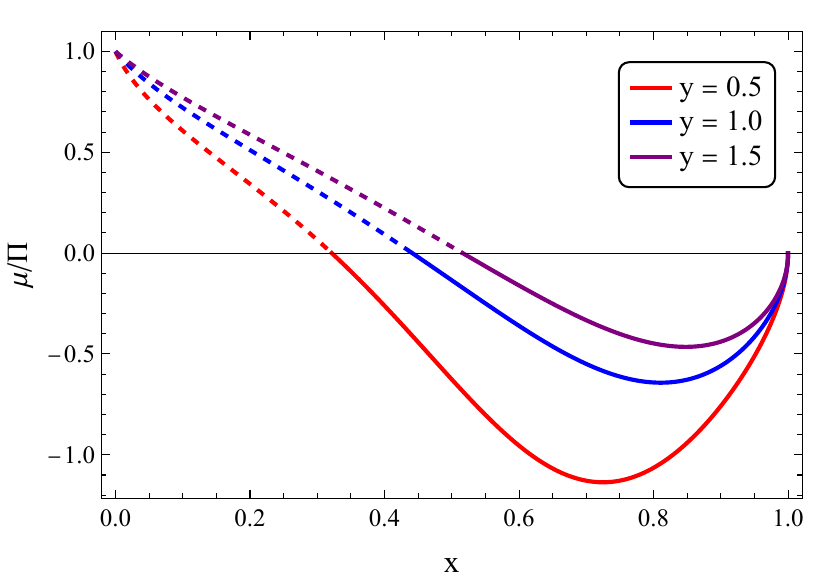}\label{subfig:Eos2}}
\caption{Plot of $\mu(a_0)/\Pi(a_0)$ as a function of $x$ for various values of $y$.  The left plot shows $8\pi R^2\rho_{0}=0.5$, while the right plot displays $8\pi R^2 \rho_{0}=0.9$.  The solid lines denote positive $\mu(a_0)$, whereas the dashed lines indicate negative $\mu(a_0)$ (more details are provided in the text).}
\label{fig:Eos}
\end{figure}

\noindent Next, we consider the behaviour of the ratio of $\mu(a_0)$ and $\Pi(a_0)$ as the equation of state parameter $(\mu/\Pi)$ of the matter in the thin shell. This is shown in Figure \ref{fig:Eos}.
The plots in Figure \ref{fig:Eos} are constructed for two chosen allowed values of $8\pi R^2\rho_{0}$. In each plot, the different coloured lines represent $\mu/\Pi$ as a function of $x$ for different selected values of $y$. As $\Pi(a_0)$ is always negative for any parameter value, the solid-coloured lines represent the $\mu/\Pi$ where $\mu$ is positive. However, the dashed lines are for negative $\mu$ values which are not
of interest to us here.
It may be noted that for $a_0>>2M$, $\mu/\Pi$ approaches unity for any value of $y$. On the other hand, when $a_0\to2M$, $\mu/\Pi$ reaches zero value, reflecting the infinite value of $\Pi(a_0)$ as $a_0\to 2M$.
In addition, a stiff matter shell $(\mu/\Pi=-1)$ may be constructed numerically for specific parameter values.
In summary, we have now clearly shown how the regular defect geometry may be used to model a thin-shell compact object. The matter content of the thin shell may be characterized by different equations of state based on the chosen parameter values.  
Although we do not have an analytical expression for the equation of state, we have a physically reasonable equation of state that can model compact objects. Note that we chose $V(a)=0$ for simplicity. One may consider other stable potentials and explore their 
corresponding equation of state. 

\noindent Thus, through the above construction of
a possible stellar model, we have been able to illustrate (atleast theoretically) the utility of the regular defect spacetime derived in Section \ref{IV}.


\section{Reconstruction of a subclass of solutions via TOV equation}\label{V}
\noindent In this penultimate section, we regenerate a subclass of the above-discussed regular solutions from the Tolman–Oppenheimer–Volkoff (TOV) equation for an anisotropic fluid with a chosen equation of state. For this purpose, we start with the static, spherically symmetric Schwarzschild-gauge geometry mentioned in Eq.(\ref{2.3}). The energy-momentum tensor of an anisotropic fluid is also mentioned in Eq.(\ref{2.2}). The well-known TOV equation with differing radial and tangential pressures can be written as
\begin{equation}
    \frac{dp_r}{dr}=-\frac{(\rho+p_r)(m(r)+4\pi r^3 p_r)}{r(r-2m(r))}+\frac{2}{r}(p_t-p_r)
\end{equation}
where, the mass function $m(r)$ was introduced earlier through the metric function $f(r)$. To construct regular spacetime solutions, we assume the following equations of state,
\begin{equation}
    p_r=-\rho
\end{equation}
\begin{equation}\label{6.3}
    p_t=a\rho+\frac{b}{\rho_{0}^{\lambda-1}}\rho^{\lambda}
\end{equation}
Here $\rho_{0}$ is the density at $r=0$ and $a$, $b$ are dimensionless parameters. The Schwarzschild gauge symmetry of all the regular spacetimes discussed in the previous sections motivates us to assume radial pressure $(p_r)$ as the negative of the energy density $(\rho)$. The first term in Eq.(\ref{6.3}) represents the linear equation of state, while the subsequent term corresponds to a polytropic equation of state. Both of them are widely used independently in various realms of cosmology and astrophysics \cite{Clayton, Sekhmani1}. We include both of them 
in a linear combination
in our expression for $p$  with the aim of constructing regular spacetimes. In recent times, researchers have worked on this type of equation of state \cite{Luongo}.  

\noindent Substituting the above equation of state in the TOV equation, we have the following energy density distribution,
\begin{equation}
    \rho=\left(\frac{(1+a)}{1-\frac{b}{\rho_{0}^{\lambda-1}}\left(\frac{C^2}{r^2}\right)^{(1+a)(\lambda-1)}}\right)^{\frac{1}{\lambda-1}}\left(\frac{C^2}{r^2}\right)^{(1+a)}
\end{equation}
where $C^2$ is the integration constant, and it can be chosen as $C^{2(1+a)(\lambda-1)}=\rho_{0}^{\lambda-1}R^{2(1+a)(\lambda-1)}/(-b)$. Therefore, the energy density becomes,
\begin{equation}\label{6.5}
    \rho=\rho_{0}\left(-\frac{(1+a)/b}{1+\left(\frac{r}{R}\right)^{2(1+a)(\lambda-1)}}\right)^{\frac{1}{\lambda-1}}
\end{equation}
One may note that the above energy density can represent a subclass of the parametrised Dekel-Zhao density profile (\ref{2.5}) for certain parameter values. If we consider $b=-(\alpha + 3)/2$, $a=(\alpha+1)/2$ and $\lambda=\frac{\nu}{\alpha+3}+1$ the above density profile
corresponds to the $\mu=3$ profile of the parametrised density distribution introduced at the beginning of this article.

\noindent {\sf{Allowed values of model parameters :}}
\noindent In principle, the model parameters of the above density profile can take any real value, and by integrating Einstein's equations, we have the corresponding mass function $m(r)$. However, to construct a real physical spacetime, we impose the following restrictions,

\noindent (i) The spacetime solution should be asymptotically flat, i.e. as $r\to \infty$, $\rho$, $p_r$ and $p_t$ should vanish, which leads
\begin{equation}
    a+1 > 0, \hspace{1cm} \lambda>0
\end{equation}

\noindent (ii) The requirement of real and positive energy density implies
\begin{equation}
    b<0
\end{equation}

\begin{figure}[h]
\centering
\includegraphics[width=0.6\textwidth]{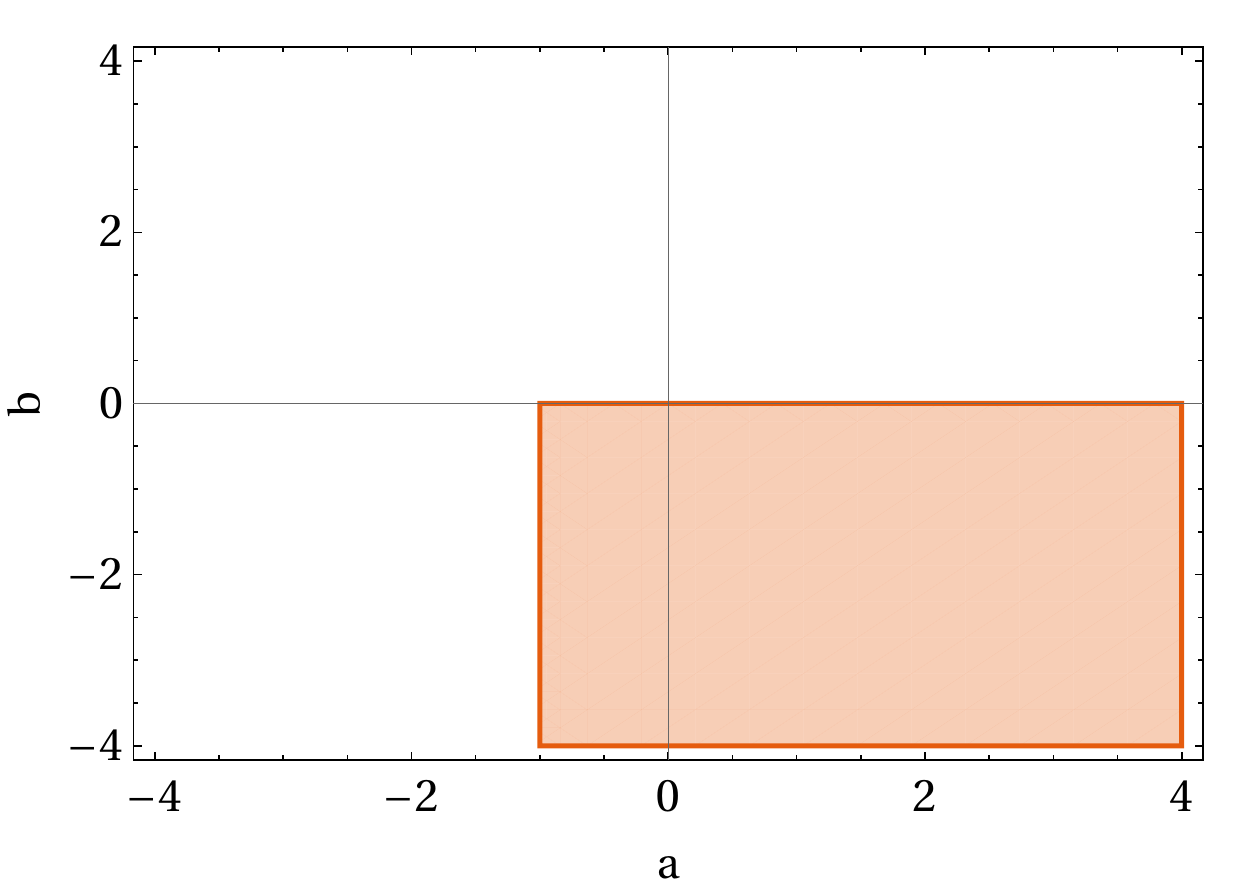}
\caption{Plot representing the allowed domain of the model parameter space.}
\label{fig:Allowedregion}
\end{figure}

\noindent Therefore, we are bound to work in a regulated region of the parameter space $a,b$, as illustrated in Figure \ref{fig:Allowedregion}.

\noindent {\sf{Regular solutions from the allowed parameter values :}}
\noindent We now choose some specific allowed parameter values ($a,b,\lambda$) that may be used to reconstruct the regular spacetimes discussed in earlier sections. The procedure for the reconstruction of solutions is as follows: (a) we have the density profile from Eq.(\ref{6.5}) for certain parameter values, (b) from the equation of state, the other components of the energy-momentum tensor can be obtained, (c) hence, by solving Einstein equations, we have the mass function $m(r)$.

\noindent (i) When $a=\frac{1}{2}$, $b=-\frac{3}{2}$ and $\lambda=\frac{5}{3}$, the density can be identified with the King dark matter density profile \cite{King} in Eq.(\ref{3.1}). As a result, from Einstein equation, we reconstruct our new regular black hole solution with the following metric function (mentioned earlier) 
\begin{equation}
    f(r)=1+\frac{8\pi \rho_{0}R^3}{\sqrt{r^2+R^2}}+\frac{8\pi \rho_{0}R^3}{r}\ln{\left(\frac{\sqrt{r^2+R^2}-r}{R}\right)}
\end{equation}
\\
\noindent (ii) The pseudo-isothermal dark matter density profile \cite{Isothermal} is obtained for $a=0$, $b=-1$, and $\lambda=2$. Thus, we recover the regular defect geometry from the Einstein equations with the following metric function, 
\begin{equation}
    f(r)=1-8\pi R^2\rho_{0}+8\pi R^2\rho_{0}\frac{\arctan(r/R)}{r/R} 
\end{equation}
The other known solutions are as follows:
\\
\noindent (iii) We have the Dymnikova regular black hole solution when $a=1$, $b=-2$ and $\lambda=\frac{3}{2}$.
\\
\noindent (iv) The Bronnikov regular solution can be obtained for $a=1$, $b=-2$ and $\lambda=\frac{7}{4}$.
\\
\noindent (v) $a=\frac{3}{2}$, $b=-\frac{5}{2}$ and $\lambda=\frac{7}{5}$ leads to Bardeen regular black hole.
\\
\noindent (vi) $a=2$, $b=-3$ and $\lambda=\frac{3}{2}$ represents Hayward solution.

\noindent Therefore, we are able rederive the regular solutions discussed earlier through the TOV equation approach. It is possible to explore other permissible values of model parameters and construct 
their corresponding solution. In our work here, we have discussed an equation of state that characterises a subclass of the parametrised Dekel-Zhao density profile. A more general equation of state for matter, which can describe the full density profile, may perhaps be of
greater significance. This remains an open issue.


\section{Summary and concluding remarks}\label{VII}
\noindent Let us now summarise our work and conclude with a few 
remarks.

\noindent (i) We use the Dekel-Zhao parametrised density profile 
which represents several known dark matter density profiles for specific parameter choices. In the framework of general relativity, the corresponding spacetime solutions in the Schwarzschild gauge, sourced by such density profiles (for specific parameter values), are found to be regular. The related pressure profiles are also found
as a consequence. The use of the dark matter density profiles
as a choice for the density in our models, do not have any direct physical link with cold dark matter models, which have negligible pressures.

\noindent (ii) We have found a new regular black hole solution when the energy density of the required matter is identified with King dark matter density profile. The regularity of the geometry is confirmed by analysing the curvature scalars and geodesic completeness. The required matter energy-momentum tensor obeys NEC and WEC. However, SEC does not hold when $r<\sqrt{2}R$. We model the source of the geometry as a magnetic monopole governed by a specific nonlinear electrodynamics model. The shadow radius of this regular black hole is also computed and does tally with available observational results from EHT. 
Moreover, we show that some of the known regular black holes can 
also be reconstructed from the Dekel-Zhao parametrised density profile for chosen parameter values.

\noindent (iii) A subclass of the parametrised density profile is found, which can source new spacetimes with defects. For a specific parameter choice which leads to the pseudo-isothermal density profile, we have a new defect solution free from any curvature singularity. It is found that a polytropic fluid of strings can source our regular defect geometry.

\noindent (iv) Next, we use the regular defect geometry to construct a dynamically stable star, similar to a gravastar model. The interior defect geometry is joined to an outer Schwarzschild geometry through a thin shell. The surface energy density and surface tension of the thin shell are found. We obtain the equation of state for the matter in the thin shell as a function of the metric parameters.

\noindent (v) Finally, we reconstruct the regular solutions discussed in this paper from the anisotropic TOV equation with a chosen equation of state. We observe that the density profile constructed from the TOV equation is a subclass of the parametrised Dekel-Zhao density profile. In Table \ref{table2}, we summarise the density profiles and the corresponding spacetime solutions.
\begin{table}[h]
\centering
\begin{tabular}{ |c|c| } 
\hline
{Density distribution $\rho(r)$} & Regular metric $f(r)$ \\
\hline
 $\begin{array}{c}\rho_{0}\left(1+\frac{r^2}{R^2}\right)^{-3/2} \\ \text{King profile} \end{array}$ & $1+\frac{8\pi \rho_{0}R^{3}}{\sqrt{r^2+R^2}}+\frac{8\pi \rho_{0}R^3}{r}\log\left(\frac{\sqrt{r^2+R^2}-r}{R}\right)$\\ 
\hline
 $\begin{array}{c}\rho_{0}\left(1+\frac{r^2}{R^2}\right)^{-1} \\ \text{Pseudo Isothermal profile} \end{array}$ & $1-8\pi R^2\rho_{0}+8\pi R^2\rho_{0}\frac{\tan^{-1}\left(\frac{r}{R}\right)}{\frac{r}{R}}$ \\
\hline
 $\rho_{0}\left(1+\frac{r^2}{R^2}\right)^{-2}$ & $\begin{array}{c}1-8\pi R^2\rho_{0}\frac{\tan^{-1}(r/R)}{2r/R}+\frac{4\pi\rho_{0}R^{4}}{(r^2+R^2)} \\ \text{Dymnikova solution}\end{array}$  \\
\hline
$\rho_{0}\left(1+\frac{r^3}{R^3}\right)^{-\frac{4}{3}}$ & $\begin{array}{c}f(r)=1-\frac{8\pi\rho_{0}R^3}{r}\left\{1-\left(1+\left(\frac{r}{R}\right)^{3}\right)^{-\frac{1}{3}}\right\} \\ \text{Bronnikov solution}\end{array}$  \\
\hline
$\begin{array}{c}\rho_{0}\left(1+\frac{r^2}{R^2}\right)^{-5/2} \\ \text{Plummer profile}\end{array}$ & $\begin{array}{c}1-\frac{8\pi\rho_{0}R^3 r^2}{3(r^2+R^2)^{3/2}}\\ \text{Bardeen solution}\end{array}$\\
\hline
 $\rho_{0}\left(1+\frac{r^3}{R^3}\right)^{-2}$ & $\begin{array}{c}1-\frac{8\pi\rho_{0}R^3 r^2}{3(r^3+R^3)}\\ \text{Hayward solution}\end{array}$\\
\hline
\end{tabular}
\caption{Summary of the density distributions and regular solutions.}
\label{table2}
\end{table}

\

\noindent In conclusion, we have considered the parametrised Dekel-Zhao density profile which may be used to generate various regular spacetime solutions (in GR). We have worked on some specific regular black holes and regular defect solutions which arise for certain parameter values.  Other 
choices for the parameters may still be explored in a similar way to construct newer regular solutions (black holes or defects). Moreover, one may, through the TOV equation and an equation of state different from the
expression in Eq.(\ref{6.3}), construct other parametrised density profiles 
yielding additional examples of regular spacetimes. 

\noindent Since we have new regular spacetime geometries, it will be
interesting to find out possible observational signatures. Here, we have
done it briefly for the regular black hole, by calculating the shadow profile and, for the defect spacetime, by theoretically constructing a possible stellar model. One can surely investigate gravitational lensing, time delay
as well as different types of orbits in general (through a study of geodesics) for our regular spacetimes. Another direction for future work is related to the construction of rotating versions of the static, spherically symmetric
line elements obtained here. Knowing the rotating versions is surely an important step towards developing astrophysically relevant signatures
in different scenarios.  Finally, one may also find the quasinormal modes of perturbations (scalar and gravitational) which can be of use
while considering these geometries (and their rotating versions) as black hole mimickers in the context of gravitational wave astronomy.

\section*{Acknowledgements}
\noindent 
AK is grateful to Pritam Banerjee, Poulami Dutta Roy and Soumya Jana for their valuable feedback. He also thanks the Indian Institute of Technology Kharagpur, India, for support through a fellowship.


\begin{thebibliography}{Rubinsteinetal}

\bibitem{penrose} R. Penrose, \href{https://doi.org/10.1103/PhysRevLett.14.57}{Phys. Rev. Lett. {\bf 14}, 57 (1965).}

\bibitem{hawking} S. W. Hawking and R. Penrose, \href{https://doi.org/10.1098/rspa.1970.0021}{Proc. Roy.
Soc. Lond. {\bf A 314}, 529 (1970).}

\bibitem{senovilla} J. M. M. Senovilla and D. Garfinkle, \href{http://dx.doi.org/10.1088/0264-9381/32/12/124008}{Class. Quant.
Grav. {\bf 32}, 124008 (2015).}

\bibitem{Horowitz} G. T. Horowitz and D. Marolf, \href{https://doi.org/10.1103/PhysRevD.52.5670}{Phys. Rev. D \textbf{52}, 5670
(1995).}

\bibitem{Hofmann} S. Hofmann and M. Schneider, \href{https://doi.org/10.1103/PhysRevD.91.125028}{Phys. Rev. D \textbf{91}, 125028
(2015).}

\bibitem{Casals} M. Casals, A. Fabbri, C. Martínez, and J. Zanelli, \href{https://doi.org/10.1103/PhysRevLett.118.131102}{Phys. Rev. Lett. \textbf{118}, 131102 (2017).}

\bibitem{borissova} J. Borissova, S. Liberati and M. Visser, 
\href{https://doi.org/10.48550/arXiv.2502.00548}{arXiv:2502.00548 [gr-qc].}

\bibitem{Ayon1} E. Ayon-Beato and A. Garcia, \href{https://doi.org/10.1103/PhysRevLett.80.5056}{Phys. Rev. Lett. \textbf{80}, 5056 (1998).}

\bibitem{Ayon2}  E. Ayon-Beato and A. Garcia, \href{https://doi.org/10.1016/S0370-2693%2899%2901038-2}{Phys. Lett. B \textbf{464}, 25 (1999).}

\bibitem{Ayon3} E. Ayon-Beato and A. Garcia, \href{https://doi.org/10.1023/A:1026640911319}{Gen. Relativ. Gravit. \textbf{31}, 629 (1999).}

\bibitem{Ayon4} E. Ayon-Beato and A. Garcia, \href{https://doi.org/10.1016/S0370-2693%2800%2901125-4}{Phys. Lett. B \textbf{493}, 149 (2000).}

\bibitem{Ayon5} E. Ayon-Beato and A. Garcia, \href{https://doi.org/10.1007/s10714-005-0050-y}{Gen. Relativ. Gravit. \textbf{37}, 635 (2005).}

\bibitem{Bronnikov1}K. A. Bronnikov, \href{https://doi.org/10.1142/S0218271818410055}{Int. J. Mod. Phys. D \textbf{27}, 1841005 (2018).}

\bibitem{Balart} L. Balart and E. C. Vagenas, \href{https://link.aps.org/doi/10.1103/PhysRevD.90.124045}{Phys. Rev. D \textbf{90}, 124045 (2014).}

\bibitem{Bronnikov3} K. Bronnikov, \href{https://doi.org/10.1103/PhysRevD.63.044005}{Phys. Rev. D \textbf{63}, 044005 (2001).}

\bibitem{Poshteh} M. B. Jahani Poshteh and N. Riazi, \href{https://doi.org/10.1142/S0218271821500796}{Int. J. Mod. Phys. D 30,
2150079 (2021).}

\bibitem{Kar1} A. Kar and S. Kar, \href{ https://doi.org/10.1007/s10714-024-03238-4}{Gen. Relativ. Gravit. \textbf{56}, 52 (2024).}

\bibitem{Kar2} A. Kar, \href{https://doi.org/10.1140/epjc/s10052-024-13603-x}{Eur. Phys. J. C \textbf{84}, 1246 (2024).}

\bibitem{Bronnikov2} K. A. Bronnikov and J. C Fabris, \href{ https://doi.org/10.1103/PhysRevLett.96.251101}{Phys. Rev. {\bf D 96}, 251101 (2006)}

\bibitem{bronnikov-rev} K. A Bronnikov,
\href{https://doi.org/10.3390/particles1010005}{ Particles {\bf 1} 56,(2018).}

\bibitem{ovalle} J. Ovalle, R. Casadio and A. Giusti, \href{https://doi.org/10.1016/j.physletb.2023.138085}{Phys. Lett. B \textbf{844}, 138085 (2023). }

\bibitem{Bueno} P. Bueno, P. A. Cano and R. A. Hennigar, \href{https://doi.org/10.48550/arXiv.2403.04827}{arXiv:2403.04827 [gr-qc]}

\bibitem{khlopov} I. Dymnikova and M. Khlopov, \href{https://doi.org/10.1142/S0218271815450029}{Int. Jr. Mod. Phys. {\bf D 24}, 1545002 (2015).}

\bibitem{zhao} H. Zhao, \href{https://doi.org/10.1093/mnras/278.2.488}{Mon. Not. Roy. Astron. Soc. {\bf 278}, 488 (1996).}

\bibitem{zhao2} H. Zhao, \href{https://doi.org/10.1093/mnras/287.3.525}{Mon. Not. Roy. Astron. Soc. {\bf 287}, 525 (1997).}

\bibitem{dekel} A. Dekel, G. Ishai, A. A. Dutton, A. V. Maccio, \href{https://doi.org/10.1093/mnras/stx486}{Mon. Not. Roy. Astron. Soc. {\bf 468}, 1005 (2017).}

\bibitem{dz1} J. Freundlich, F. Jiang, A. Dekel, N. Cornuault, O. Ginzburg, R. Koskas, S. Lapiner, A. Dutton, A. V. Macciò, \href{https://doi.org/10.1093/mnras/staa2790}{Mon. Not. Roy. Astron. Soc. {\bf 499 }, 2912 (2020).}

\bibitem{Salucci} P. Salucci, \href{https://doi.org/10.3390/universe11020067}{Universe \textbf{11}, 67 (2025)}

\bibitem{dz2} D. Batic, J. M. Faraji, M. Nowakowski, \href{https://doi.org/10.1140/epjc/s10052-022-10731-0}{Eur. Phys. J. {\bf C 82}, 759 (2022).}

\bibitem{Sekhmani2} A. A. Badawi, S. Shaymatov and Y. Sekhmani, \href{https://doi.org/10.1088/1475-7516/2025/02/014}{JCAP \textbf{02}, 014 (2025).}

\bibitem{ovgun} A. Ovgun and R. C. Pantig, \href{https://doi.org/10.48550/arXiv.2501.12559}{arXiv 2501.12559 [gr-qc].}

\bibitem{Barriola} M. Barriola and A. Vilenkin, \href{https://doi.org/10.1103/PhysRevLett.63.341}{Phys. Rev. Lett. \textbf{63}, 341 (1989).}

\bibitem{Nucamendi} U. Nucamendi, M. Salgado and D. Sudarsky, \href{https://doi.org/10.1103/PhysRevLett.84.3037}{Phys. Rev. Lett. \textbf{84}, 3037 (2000).}

\bibitem{Marunovic} A. Marunovi\'c and M. Murkovi\'c, \href{https://doi.org/10.1088/0264-9381/31/4/045010}{Class. Quantum Grav. \textbf{31}, 045010 (2014).} 

\bibitem{Carames} T. R. P. Caram\^es, \href{https://doi.org/10.1103/PhysRevD.108.084002}{Phys. Rev. D \textbf{108}, 084002 (2023).}


\bibitem{Vilenkin} A. Vilenkin and E. P. S. Shellard, {\em Cosmic Strings and Other Topological Defects} (Cambridge University Press, Cambridge, England, 2000).


\bibitem{Luongo} S. N. Sajadi, S. Ponglertsakul and O. Luongo, \href{https://doi.org/10.48550/arXiv.2502.02098}{arXiv.2502.02098[gr-qc].}

\bibitem{NFW} J. F. Navarro, C. S. Frenk and S. D.M. White, \href{https://doi.org/10.1086/177173}{Astrophys. J. \textbf{462}, 563 (1996).}

\bibitem{King} I. King, \href{https://doi.org/10.1086/108756}{Astron. J. \textbf{67}, 471 (1962).}

\bibitem{Isothermal} K.G. Begeman, A.H. Broeils and R.H. Sanders, \href{https://doi.org/10.1093/mnras/249.3.523}{Mon. Not. Roy. Astron. Soc. \textbf{249}, 523 (1991).}

\bibitem{Plummer} H. C. Plummer,\href{https://doi.org/10.1093/mnras/71.5.460}{Mon. Not. Roy. Astron. Soc., \textbf{71}, 460-470 (1911).}

\bibitem{Hernquist} L. Hernquist, \href{https://doi.org/10.1086/168845}{Astrophys. J. \textbf{356}, 359 (1990).}


\bibitem{Zakhary} E. Zakhary and C. B. G.  Mcintosh, \href{https://doi.org/10.1023/A:1018851201784}{Gen. Rel. Grav. {\bf 29}, 539 (1997).}

\bibitem{Hu} H. W. Hu, C. Lan and Y. G. Miao Lan,  \href{https://doi.org/10.1140/epjc/s10052-023-12228-w}{Eur. Phys. J. C \textbf{83}, 1047 (2023).}

\bibitem{Chrusciel} P. T. Chru\'sciel, M. A. MacCallum, and D. B. Singleton,  \href{https://doi.org/10.1098/rsta.1995.0004}{Phil. Trans. R. Soc. A \textbf{350}, 113–141 (1995).}

\bibitem{Godazgar} M. Godazgar and G. Macaulay, \href{https://doi.org/10.1103/PhysRevD.102.064036}{Phys. Rev. D \textbf{102}, 064036 (2020).}

\bibitem{He} X. He, X. Wu and N. Xie, \href{https://doi.org/10.48550/arXiv.2504.07774}{arXiv:2504.07774 [gr-qc].}

\bibitem{vishu} C. V. Vishveshwara, \href{https://doi.org/10.1063/1.1664717}{J. Math. Phys. {\bf 9}, 1319 (1968).}

\bibitem{mt} M. S. Morris and K. S. Thorne, \href{https://doi.org/10.1119/1.15620}{Am. Jr. Phys. {\bf 56}, 395 (1988).} 

\bibitem{Hawking} S.W. Hawking and G.F.R. Ellis, {\em The Large Scale Structure of Spacetime} (Cambridge University Press, Cambridge 1973).

\bibitem{Wald}R. M. Wald, {\em General Relativity}. Chicago Univ. Pr., Chicago, USA, 1984.

\bibitem{Modesto} T. Zhou and L. Modesto, \href{https://link.aps.org/doi/10.1103/PhysRevD.107.044016}
{Phys. Rev. D \textbf{107}, 044016 (2023).}

\bibitem{Carballo1}R. Carballo-Rubio, F. Di Filippo, S. Liberati, and M. Visser, \href{https://doi.org/10.1103/PhysRevD.101.084047}{Phys. Rev. D \textbf{101}, 084047 (2020).}

\bibitem{Carballo2}R. Carballo-Rubio, F. Di Filippo, S. Liberati, and M. Visser, \href{https://doi.org/10.1007/JHEP02(2022)122}{J. High Energ. Phys. \textbf{02}, 122 (2022).}


\bibitem{Ansoldi} S. Ansoldi, {\em in Conference on Black Holes and Naked Singularities, Milano, Italy, 2007}, \href{https://doi.org/10.48550/arXiv.0802.0330}{arXiv: 0802.0330 [gr-qc].}

\bibitem{Zaslavskii} O. B. Zaslavskii, \href{https://doi.org/10.1016/j.physletb.2010.04.031}{Phys. Lett. B  \textbf{688}, 278–280 (2010).}

\bibitem{Virbhadra} C. M. Claudel, K.S. Virbhadra, G.F.R. Ellis, \href{https://doi.org/10.1063/1.1308507}{J.Math.Phys. \textbf{42}, 818-838 (2001).}

\bibitem{bardeenshadow} J.M. Bardeen, \href{https://www.google.co.in/books/edition/Black_Holes/sUr-EVqZLckC?hl=en&gbpv=1&pg=PA215&printsec=frontcover}{ in Black Holes (Les Astres Occlus), ed. C. DeWitt and B. S. DeWitt (New York: Gordon and Breach), 215 (1973).} 

\bibitem{Volker} V. Perlick and O. Y. Tsupko, \href{https://doi.org/10.1016/j.physrep.2021.10.004}{Phys. Rep. \textbf{947}, 1-39 (2022).}

\bibitem{Chen}  D. Chen, C. Gao, X. Liu, and C. Yu, \href{https://doi.org/10.1140/epjc/s10052-021-09510-0}{Eur. Phys. J. C, \textbf{81}, 700 (2021).}

\bibitem{Akiyama1} Event Horizon Telescope Collaboration et al, \href{https://dx.doi.org/10.3847/2041-8213/ab0ec7}{Astrophys. J. Lett. \textbf{875}, L1 (2019).}

\bibitem{Akiyama5} Event Horizon Telescope Collaboration et al., \href{https://dx.doi.org/10.3847/2041-8213/ab0f43}{Astrophys. J. Lett. \textbf{875}, L5 (2019).}

\bibitem{Akiyama6} Event Horizon Telescope Collaboration et al., \href{https://dx.doi.org/10.3847/2041-8213/ab1141}{Astrophys. J. Lett. \textbf{875}, L6 (2019).}


\bibitem{Medeiros} L. Medeiros et al., \href{https://dx.doi.org/10.3847/2041-8213/acc32d}{Astrophys. J. Lett. \textbf{947}, L7 (2023).}

\bibitem{Blakeslee} J. P. Blakeslee et al., \href{https://dx.doi.org/10.1088/0004-637X/694/1/556}{Astrophys. J. Lett. \textbf{694}, 556–572 (2009).}

\bibitem{Bird} S. Bird, W. E. Harris, J. P. Blakeslee, and C. Flynn, \href{https://doi.org/10.1051/0004-6361/201014876}{Astron. Astrophys. \textbf{524}, A71 (2010).}

\bibitem{Cantiello} M. Cantiello et al., \href{https://dx.doi.org/10.3847/2041-8213/aaad64}{Astrophys. J. Lett. \textbf{854}, L31 (2018).}

\bibitem{Akiyama12} Event Horizon Telescope Collaboration et al., \href{https://dx.doi.org/10.3847/2041-8213/ac6674}{Astrophys. J. Lett. \textbf{930}, L12 (2022).}

\bibitem{Akiyama17} Event Horizon Telescope Collaboration et al., \href{https://dx.doi.org/10.3847/2041-8213/ac6756}{Astrophys. J. Lett. \textbf{930} L17 (2022).}

\bibitem{Do} T. Do et al., \href{https://doi.org/10.1126/science.aav8137}{Science \textbf{365} no. 6454, 664–668 (2019).}

\bibitem{Gravity1} R. Abuter et al., \href{https://doi.org/10.1051/0004-6361/202142465}{Astron. Astrophys. \textbf{657}, L12 (2022).}

\bibitem{Gravity2} R. Abuter et al., \href{https://doi.org/10.1051/0004-6361/202037813}{Astron. Astrophys. \textbf{636}, L5 (2020).}

\bibitem{Dymnikova} I. Dymnikova, \href{https://doi.org/10.1088/0264-9381/21/18/009}{Class. Quant. Grav. \textbf{21}, 4417-4429 (2004)}.

\bibitem{Fan} Z. Y. Fan and X. Wang, \href{https://doi.org/10.1103/PhysRevD.94.124027}{Phys. Rev. D \textbf{94}, 124027 (2016).}

\bibitem{Neves} J.C.S. Neves and A. Saa, \href{https://doi.org/10.1016/j.physletb.2014.05.026}{Phys. Lett. B \textbf{734}, 44 (2014).}

\bibitem{Bardeen} J. M. Bardeen, {\em in Proceedings of International Conference GR5, Tbilisi, USSR, 1968}, p. 174.

\bibitem{Hayward} S. A. Hayward, \href{https://link.aps.org/doi/10.1103/PhysRevLett.96.031103}{Phys. Rev. Lett. \textbf{96}, 031103 (2006).}

\bibitem{Letelier1} P. S. Letelier, \href{https://link.aps.org/doi/10.1103/PhysRevD.20.1294}{Phys. Rev. D \textbf{20}, 1294 (1979).}

\bibitem{Letelier2} P. S. Letelier, \href{https://doi.org/10.1007/BF02755096}{Nuov. Cim. B \textbf{63}, 519–528 (1981).}

\bibitem{Soleng} H. H Soleng, \href{https://doi.org/10.1007/BF02107935}{Gen. Rel. Grav. \textbf{27}, 367–378 (1995).}

\bibitem{Santos} L. C. N. d. Santos, \href{https://doi.org/10.48550/arXiv.2502.15846}{arXiv:2502.15846 [gr-qc].}

\bibitem{sen} N. Sen, Annalen der Physik 378, 365 (1924).

\bibitem{Darmois} G. Darmois, Memorial Des Sciences Mathematiques (Gauthier-Villars, Paris, 1927).

\bibitem{Israel} W. Israel, \href{https://doi.org/10.1007/BF02712210}{Nuovo Cimento B (1965-1970) \textbf{48}, 463 (1967).}

\bibitem{Visser1} M. Visser and D.L. Wiltshire, \href{https://doi.org/10.1088/0264-9381/21/4/027}{Class. Quantum Gravit. \textbf{21}, 1135 (2004).}

\bibitem{Visser2}  M. Visser, \href{https://doi.org/10.1103/PhysRevD.39.3182}{Phys. Rev. D \textbf{39}, 3182 (1989).}

\bibitem{Visser3} M. Visser, {\em Lorentzian Wormholes: From Einstein to Hawking} (AIP, Woodbury, p. 412, 1995).

\bibitem{poisson} E. Poisson, {\em A Relativist's Toolkit: the mathematics of black hole mechanics}, Cambridge University Press, Cambridge, UK (2004).

\bibitem{Visser5} P. M. Moruno, N. M. Garcia, F. S. N. Lobo and M. Visser, \href{https://doi.org/10.1088/1475-7516/2012/03/034}{JCAP \textbf{03}, 034 (2012).}

\bibitem{Visser4} N. M. Garcia, F. S. N. Lobo and M. Visser, \href{https://doi.org/10.1103/PhysRevD.86.044026}{Phys. Rev. D \textbf{86}, 044026 (2012).}

\bibitem{Lobo} F. S. N. Lobo and P. Crawford, \href{https://doi.org/10.1088/0264-9381/22/22/012}{Class. Quantum Grav. \textbf{22}, 4869 (2005).}

\bibitem{Bhattacharjee} D. Bhattacharjee and P. K. Chattopadhyay, \href{https://doi.org/10.1016/j.jheap.2024.08.001}{J. High Energ. Astrophys. \textbf{43}, 248-257 (2024).}

\bibitem{Lanczos} K. Lanczos, \href{ https://doi.org/10.1002/andp.19243791403}{Ann. Phys. \textbf{379}, 518 (1924).}


\bibitem{Clayton} D.D. Clayton, {\em Principles of Stellar Evolution and Nucleosynthesis} (McGraw-Hill, 1968).

\bibitem{Sekhmani1} Y. Sekhmani, S. Zare, L. Nieto, H. Hassanabadi, and K. Boshkayev, \href{https://doi.org/10.1016/j.jheap.2025.100389}{J. High Energ. Astrophys. \textbf{47}, 100389 (2025).} 



\end{thebibliography}
\end{document}